\documentclass[article,twocolumn]{revtex4}

\usepackage{graphicx}
\usepackage{dcolumn}
\usepackage{bm}
\usepackage{color}

\newcommand{\etal}{{\it et al.}}
\begin{document}


\title[Comparison of different methods for analyzing $\mu$SR line shapes]{Comparison
of different methods for analyzing $\mu$SR line shapes in the
vortex state of type-II superconductors}

\author{A.~Maisuradze}
\email{alexander.maisuradze@psi.ch}
\affiliation{Physik-Institut der Universit\"{a}t Z\"{u}rich, Winterthurerstrasse 190, CH-8057 Z\"{u}rich, Switzerland}
\affiliation{Laboratory for Muon Spin Spectroscopy, Paul Scherrer Institut, CH-5232 Villigen PSI, Switzerland}
\author{R.~Khasanov}
\email{rustem.khasanov@psi.ch}
\affiliation{Physik-Institut der Universit\"{a}t Z\"{u}rich, Winterthurerstrasse 190, CH-8057 Z\"{u}rich, Switzerland}
\affiliation{Laboratory for Muon Spin Spectroscopy, Paul Scherrer Institut, CH-5232 Villigen PSI, Switzerland}
\author{A.~Shengelaya}
\affiliation{Physics Institute of Tbilisi State University, Chavchavadze 3, GE-0128 Tbilisi, Georgia}
\author{H.~Keller}
\affiliation{Physik-Institut der Universit\"{a}t Z\"{u}rich, Winterthurerstrasse 190, CH-8057 Z\"{u}rich, Switzerland}

\begin{abstract}

A detailed analysis of muon-spin rotation ($\mu$SR) spectra in the vortex state of type-II superconductors using different theoretical models is presented. Analytical approximations of the London and Ginzburg-Landau (GL) models, as well as an exact solution of the GL model were used. The limits of the validity of these models and the reliability to extract parameters such as the magnetic penetration depth $\lambda$ and the coherence length $\xi$ from the experimental $\mu$SR spectra were investigated.
The analysis of the simulated $\mu$SR spectra showed that at  high magnetic fields there is a strong correlation between obtained  $\lambda$ and $\xi$ for any value of the Ginzburg-Landau parameter $\kappa = \lambda/\xi$. The smaller the applied magnetic field is, the smaller is the possibility to find the correct value of $\xi$. A simultaneous determination of $\lambda$ and $\xi$ without any restrictions is very problematic, independent of the model used to describe the vortex state.
It was found that for extreme type-II superconductors and low magnetic fields, the fitted value of $\lambda$ is  practically independent of $\xi$.
The second-moment method frequently used to analyze $\mu$SR spectra by means of a multi-component Gaussian fit, generally yields reliable values of $\lambda$ in the whole range of applied fields $ H_{c1} \ll H \lesssim H_{c2}$ ($H_{c1}$ and $H_{c2}$ are the first and second critical fields, respectively).
These results are also relevant for the interpretation of small-angle neutron scattering (SANS) experiments of the vortex state in type-II superconductors.
\end{abstract}
%
%
\maketitle

\section{Introduction}

The muon-spin rotation ($\mu$SR) technique is one of the most powerful and unique tool to study the internal magnetic field distribution $P(B)$ associated with the vortex lattice in type-II superconductors (see, e.g. Refs. \onlinecite{ShenkBook, Reotier97review, Sonier00RMP}). In the vortex state for an applied magnetic field $H > H_{c1}$, or $B>0$ ($H_{c1}$ and $B$ are the first critical field and the magnetic induction in a sample, respectively) \cite{Brandt03} the energy of the surface, separating normal and superconducting fractions of the sample becomes negative and the field penetrates the sample
in the form of quantized flux lines, called vortices each of them
containing an elementary flux quantum ($\Phi_0=h/2e \simeq 2.0678\cdot 10^{-15}$~Wb) \cite{Abricosov50}.
In the case of small pinning these vortices arrange themselves in a regular vortex lattice called a flux-line lattice (FLL) \cite{Abricosov50}. The distribution of the internal magnetic fields $P(B)$ inside the
superconducting sample in the vortex state is uniquely determined by two characteristic
lengths, the magnetic field penetration depth $\lambda$ and the
coherence length $\xi$.
From $\mu$SR experiments $P(B)$ profiles are obtained by performing a Fourier transformation of the $\mu$SR time spectra.
There are different approaches to analyze $\mu$SR data. Generally, the magnetic field penetration depth $\lambda$ is determined from the second moment $\langle\Delta B^2\rangle$ of
the internal field distribution $P(B)$ \cite{Aeppli87, Harshman87, Uemura88,  Pumpin90, Zimmermann95, Aegerter97, Lee99}.
For an isotropic extreme
type-II superconductor ($\lambda\gg\xi$) it was shown that
$\langle\Delta B^2\rangle\propto\lambda^{-4}$ \cite{Brandt88PRB}.
The more advanced approaches that allow to obtain not only
$\lambda$, but also the coherence length $\xi$, require a
theoretical model for the spatial variation of the internal
magnetic field $B({\bf r})$ (${\bf r}$ is spatial coordinate). An essential requirement of the model
is that it must account for the finite size of the vortex cores. So
far, the internal magnetic field distribution $P(B)$ measured by
$\mu$SR was analyzed assuming analytical models for $B({\bf r})$
based on London and Ginzburg-Landau (GL) theories.
The London theory provides the simplest approach to model the FLL. Since London theory does not account for the finite size of
the vortex cores, a cutoff factor derived from GL theory must be
inserted into the analytical London expression for $B({\bf r})$ to
correct for the divergence of $B({\bf r})$ in the vortex core \cite{Yaouanc97,Sonier00RMP}.
The GL theory has the spatial dependence of the order
parameter build in and thus provides a phenomenological description
of the magnetic field profile in the vortex core region. Abrikosov \cite{Abricosov50} predicted the vortex core state from a periodic solution of the GL equations near the second critical field $B_{c2} = \mu_0 H_{c2}$ and provided an approximate analytical solution of these equations for an isolated vortex for fields of the
order of $H_{c1}$.  Clem \cite{Clem75} proposed a variational method
to solve the GL equations that was further extended by Hao {\it et
al.} \cite{HaoClem91}. A simplified version of this model for
$\lambda/\xi \gg 1$ was developed by Yaouanc {\it et al.} \cite{Yaouanc97}, and is often used in the literature \cite{Sonier00RMP}.

The London and the GL models
were widely applied to determine values of $\lambda$
and $\xi$ from measured $\mu$SR time spectra taken in the mixed state of type-II superconductors \cite{Riesman95,Yaouanc97,Yamashita97,Sonier00RMP,Miller00,Ohishi00,Kadono01,Price02,Ohishi02, Miller03,Sonier04,Serventi04,Sonier04JPCM,Callaghan05,Laulajainen06,Salman07,Sonier07}. We should
emphasize, however, that despite of the broad usage, the
limits of validity of these models  and the reliability of the
parameters extracted from the fits are not much discussed in the literature.
The main purpose of the present paper is to address these basic questions.
The paper is divided into two parts.
In the first part we briefly describe the models often used for the analysis of $\mu$SR spectra:
 The London model with Gaussian cutoff (LG model),  the
modified London model (ML model), and  the analytical Ginzburg-Landau
model (AGL model). These models are compared with the most precise model based on the iterative
method for solving the Ginzburg-Landau equations developed recently
by Brandt \cite{Brandt03}, the so called numerical
Ginzburg-Landau model (NGL model). $P(B)$ profiles
for various sets of $\lambda$, $\xi$, and magnetic field $B$ were
first simulated by means of the NGL model and then analyzed within the
framework of the LG, ML, and AGL model. For further discussions, it is convenient to define the reduced magnetic field $b= B/B_{c2}$.
It was found that the ML model can be used {\it only} for low
magnetic fields ($b \lesssim 0.1$), while both the AGL and the LG
model yield reliable results almost in the {\it whole} magnetic field range.
However, the values of $\lambda$ and $\xi$ obtained by means of the AGL and the
LG model deviate systematically from the initial parameters
used for the simulated $P(B)$ profiles for magnetic fields in the range
$0.01\lesssim b\leq1$. It was also shown that for $b\lesssim 0.01$
the $P(B)$ profiles do not depend on the coherence length $\xi$.
In the second part of the paper we present a systematic analysis of
simulated $\mu$SR time spectra (with typical statistics used in
real $\mu$SR experiments) by means of the LG model. In the whole field range ($0<b\leq1$) and for any values of the
Ginzburg-Landau parameter $\kappa=\lambda/\xi$ there is a strong
correlation between the values of $\lambda$ and $\xi$ determined from the
fit.  This implies that an analysis of $\mu$SR data using this approach, without taking into account these
correlations, may lead to substantial errors in the determination of the
absolute values of $\lambda$ and $\xi$, and even may result in
unphysical dependencies of $\lambda$ and $\xi$ on magnetic field
and temperature. In addition, the second moment method applied to a multiple Gaussian fit was tested in order to check how reliably the penetration depth $\lambda$ can be determined by this method. In particular, the influence of the number of Gaussians used in the multi-Gaussian fit on the quality of the fit was investigated. For typical statistics used in the experiment and practically in the whole field range ($0 < b \lesssim 1$), the second moment method applied to a multi-Gaussian fit may provide correct values for $\lambda$ within a few percent.

The paper is organized as follows: In
Sec.~II various theoretical models used to analyze $\mu$SR data are briefly described. The dependence of the
magnetic field distribution $P(B)$ on $\lambda$, $\xi$, $b$, and the
Gaussian smearing parameter $\sigma_g$, as calculated within the LG model, is discussed in
Sec.~III. In Sec.~IV we compare the results obtained by means of the models described in
Sec.~II for the case of an extreme type-II superconductor ($\kappa = \lambda/\xi \gg 1$).
Sec.~V comprises the studies of the simulated $\mu$SR data. The simulated $\mu$SR spectra
were analyzed by means of the various models described in Sec.~II in order to search for possible
correlations between the parameters, such as $\lambda$, $\xi$, and
$\sigma_g$. The conclusions follow in Sec.~VI.

\section{Models for data analysis}
As mentioned in the introduction, the simplest and
the most widely used approach for analyzing $\mu$SR data is based on the relation between  the
magnetic penetration depth $\lambda$ and the second
moment $\langle\Delta B^2\rangle$ of the internal field distribution $P_{id}(B)$ of the ideal FLL \cite{Brandt88PRB, Brandt03, LandauKeller07}:
\begin{equation}
\lambda^{-4}= C \cdot\langle\Delta B^2\rangle.
\label{eq:sec-mom_general}
\end{equation}
Here, $C$ is the proportionality coefficient depending on the value
of the reduced magnetic field $b = \langle B\rangle/B_{c2}$ [$\langle B\rangle$ is the first moment of $P_{id}(B)$] and the Ginzburg-Landau parameter $\kappa$ \cite{Brandt88PRB,Brandt03,LandauKeller07}.  In order to estimate
$\langle\Delta B^2\rangle$ one often assumes that $P_{id}(B)$ is a sum of $N$ Gaussian distributions
(generally, $N=1, 2, 3$) \cite{Weber93, Khasanov06TwoGap}:
\begin{equation}
P_{id}(B) = \frac{\gamma_\mu}{\sqrt{2 \pi}\Sigma_{i=1}^N A_i} \sum_{i=1}^N{\frac{A_i}{\sigma_i}  \exp\left[\frac{1}{2}\left(\frac{B-B_i}{\sigma_i/\gamma_{\mu}}\right)^2\right]},
\label{eq:P(B)Gauss}
\end{equation}
where $A_i$, $B_i$, and $\sigma_i/\gamma_{\mu}$ are the weight factor, the first moment, and the standard deviation of the $i-$th Gaussian component, respectively. $\gamma_{\mu} = 2\pi\times135.5342$~MHz/T is the muon gyromagnetic ratio. The first and  second moment of $P_{id}(B)$ are then readily obtained \cite{Weber93, Khasanov06TwoGap}:
\begin{equation}
\langle B \rangle =\sum_{i=1}^N{\frac{A_i\, B_i}{A_1+...+A_N}  }\, ,
\label{eq:Baver}
\end{equation}
and
\begin{equation}
\langle \Delta B^2 \rangle=\sum_{i=1}^N{\frac{A_i}{A_1+...+A_N} }
\left[ (\sigma_i/\gamma_{\mu})^2 +[B_i -  \langle B \rangle ]^2
\right] \, .
\label{eq:dB}
\end{equation}

With modern computers it became possible to develop models that allow to calculate
$P_{id}(B)$ for a FLL as a function of various parameters, such as magnetic penetration depth, coherence length, applied magnetic field,  and FLL geometry (rectangular or hexagonal) \cite{BrandtJLTPhys88and77, Riesman95, Sonier00RMP, Brandt03}. The London models (with different cut off factors) provide the simplest and fastest way to calculate $P_{id}(B)$ for the analysis of $\mu$SR data for $\kappa \gg 1$ \cite{Sonier00RMP}. Better approximations of $P_{id}(B)$ for small values of $\kappa$ and fields closer to $B_{c2}$ can be obtained by the AGL model \cite{HaoClem91, Yaouanc97}.  Strictly speaking, Ginzburg-Landau  theory is only valid in the neighborhood of the phase boundary $T_c(B)$  of a type-II superconductor. However, it is generally assumed, that Ginzburg-Landau models are also good approximations for any field and temperature.  The results obtained by the NGL model correspond to the minimum of the Ginzburg-Landau free energy, whereas other models described in this paper are just approximations to the NGL model. Therefore, the NGL model will be used as a reference for comparison with the AGL, ML, and LG models. A relatively simple method to calculate $P_{id}(B)$ within the framework of the NGL model was proposed by Brandt \cite{Brandt97_NGLmethod, Brandt03}.

In the LG, ML, AGL, and NGL approximation the spatial distribution of the magnetic field in the mixed state of a type-II superconductor is described by the Fourier expansion:
\begin{equation}\label{EqSpacialDistr}
B({\bf r}) = \langle B \rangle \sum_{\bf G}  \exp(-i{\bf Gr}) \, B_{\bf G}(\lambda, \xi) \, .
\end{equation}
Here, ${\bf r}$ is the vector coordinate in a plane perpendicular to the applied field. The origin of the coordinate system is in the center of a vortex core (see, e.g. Ref.~\onlinecite{Laulajainen06}),   ${\bf G} = 4\pi/\sqrt{3}a (m\sqrt{3}/2, n + m/2)$ are the reciprocal lattice vectors for the hexagonal FLL, $a$ is intervortex distance, $B_{\bf G}$ are the Fourier components, and $m$, $n$ are integer numbers. \\
For the LG model the Fourier components $B_{\bf G}$ are \cite{BrandtJLTPhys88and77, Sonier00RMP}:
\begin{equation}\label{EqBkLG}
B_{\bf G} = \frac{e^{-\xi^2G^2/2}  }{ 1 + G^2\lambda^2 } \; .
\end{equation}
For the ML model the Fourier components $B_{\bf G}$ are given by \cite{Riesman95, Sonier00RMP}:
\begin{equation}\label{EqBkModLond}
B_{\bf G} = \frac{e^{-\xi^2G^2/2(1-b)}  }{ 1 + G^2\lambda^2/(1-b) },
\end{equation}
For the AGL model the Fourier components $B_{\bf G}$ are \cite{HaoClem91, Yaouanc97}:
\begin{equation}\label{EqBkAGL}
B_{\bf G} = \frac{\Phi_0}{S}\frac{f_\infty K_1[\frac{\xi_v}{\lambda}(f_\infty^2 + \lambda^2G^2)^{1/2} ]}{(f_\infty^2 + \lambda^2G^2)^{1/2}K_1(\frac{\xi_v}{\lambda}f_\infty)},
\end{equation}
where $f_\infty = 1-b^4$, and
\[ \xi_v = \xi(\sqrt{2} - \frac{0.75}{\kappa})(1+b^4)^{1/2}[1 - 2b(1-b)^2]^{1/2}. \]
Here, $K_1(x)$ is the modified Bessel function. For applied magnetic fields $H\gg H_{c1}$  the relation $\mu_0 H\simeq \langle B \rangle$ holds \cite{Brandt97_NGLmethod}.
Finally, for the NGL model no analytical solution for the Fourier components $B_{\bf G}$ exists. They are determined numerically \cite{Brandt97_NGLmethod, Brandt03}.

From the known spatial distribution of the magnetic field $B({\bf r})$ in the mixed state one can extract the internal magnetic field distribution $P_{id}(B)$ for the ideal FLL by means of the following equation:
\begin{equation}\label{EqPBdistribution}
P_{id}(B) = \frac{\int\delta(B-B')dA(B')}{\int dA(B')},
\end{equation}
where $dA(B')$ is the elementary area of the FLL with a field $B'$ inside, and the integration is over a quarter of the FLL unit cell \cite{Laulajainen06}.
In order to take into account possible random deviations of the flux core positions from their ideal ones (vortex disorder) and/or possible broadening of the $\mu$SR spectra due to nuclear depolarization, one may convolute the ideal distribution $P_{id}(B)$
with a Gaussian distribution \cite{BrandtJLTPhys88and77}:
\begin{equation}\label{EqConvolution}
P(B)  = \frac{1}{\sqrt{2\pi}\sigma_g}\int P_{id}(B')\exp\left[{-\frac{1}{2}\left(\frac{B-B'}{\sigma_g}\right)^2}\right]dB' \; ,
\end{equation}
where $\sigma_g$ is the width of the Gaussian distribution. 
The relation between $\sigma_g$, vortex disorder, and nuclear depolarization is described in Sec.~\ref{sec:SigmaDepOfPB}.

The $\mu$SR time spectra can be further simulated by performing the Fourier transform of $P(B)$ convoluted with the Gaussian function given in Eq.~(\ref{EqConvolution}):
\begin{equation}\label{EqTdomainSpec}
\tilde{P}(t) = Ae^{i\phi}\int P(B)e^{i\gamma_\mu B t}dB \, ,
\end{equation}
where  $A$ and $\phi$ are the initial asymmetry and the phase of the $\mu$SR time spectra, respectively.
\begin{figure}
\center{\includegraphics[width=0.5\textwidth]{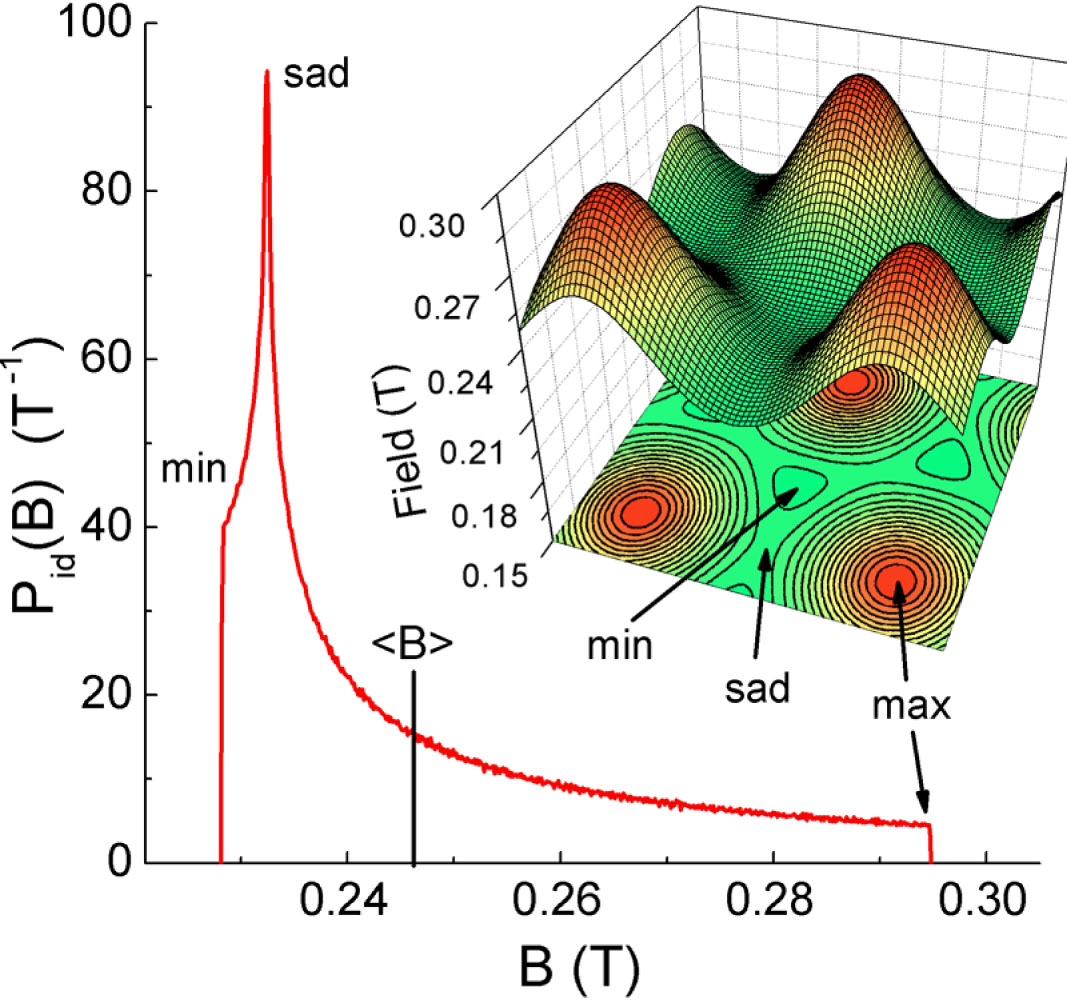}}
\caption{Example of a spatial distribution of the magnetic field $B({\bf r})$ and the corresponding local magnetic field distribution $P_{id}(B)$ for an ideal hexagonal FLL determined by the NGL method. The parameters used for
the calculations are: $\lambda = 50$~nm, $\xi = 20$~nm, and $\langle B \rangle = 0.3B_{c2} \simeq 246.8$~mT, and intervortex distance $a = 69.5$~nm. }\label{fig:ShapeDef1}
\end{figure}
For the calculations of the spatial magnetic field distribution $B({\bf r})$ in the FLL $31\times31$ Fourier components of the magnetic field and the reciprocal vector ${\bf G}$ were used. This allows to calculate the second moment of $P(B)$ with a precision of better than 10$^{-6}$. The integral in Eq.~(\ref{EqPBdistribution}) was calculated numerically over a quarter of the FLL  unit cell, divided in approximately 100x100 equal pixels, depending on the mean magnetic field $\langle B \rangle$ (see Fig.~1 of Ref.~\onlinecite{Laulajainen06}).

Equations~(\ref{EqSpacialDistr} - \ref{EqBkAGL}) are only valid for isotropic superconductors or superconductors with axial symmetry with the external magnetic field applied along the symmetry axis.  In the present study we mostly concentrate on the case of extreme type-II superconductors ($\kappa \gg 1$), such as the cuprate high-temperature superconductors. Since the qualitative behavior of $P_{id}(B)$ as a function of various parameters is essentially the same for a hexagonal and a square FFL, we will consider here only the hexagonal case.

In Fig.~\ref{fig:ShapeDef1} we plot the spatial distribution of the magnetic field $B({\bf r})$ in the mixed state and the corresponding local magnetic field distribution $P_{id}(B)$ for $\lambda = 50$~nm, $\xi  = 20$ nm, and $\langle B \rangle=0.3B_{c2} \simeq 246.8$~mT, as determined by the NGL model. The ideal FLL has three characteristic fields: (i) the maximal field $B_{max}$ corresponds to the field in the vortex core, (ii) the field at the peak of $P_{id}(B)$ is the saddle point field $B_{sad}$ (located in the middle between neighboring vortices), and (iii) the minimal field $B_{min}$ is in the center of the triangle of vortices forming the hexagonal FLL \footnote{At high fields and low temperatures the minimal and the saddle points are exchanged and the magnetic field distribution around the vortex core has a conical shape. See  Delrieu J M 1972 {\it J. Low Temp. Phys.} \textbf{6} 197; and Ref.~\onlinecite{BrandtJLTPhys88and77}.}. 
Instead of the full local magnetic field distribution $P_{id}(B)$ we will use these characteristic fields to discuss the dependence of the shape of $P_{id}(B)$ on different parameters.
\section{Dependence of $P(B)$ on $\lambda$, $\xi$, $\langle B \rangle$, and $\sigma_g$  }
In this section we concentrate on the analysis of the shape of $P(B)$ given in Eq.~(\ref{EqConvolution}) as a function of penetration depth $\lambda$ (Sec.~\ref{sec:LambdaDep}), coherence length $\xi$ (Sec.~\ref{sec:XiDep}), mean magnetic field $\langle B \rangle$ (Sec.~\ref{sec:FldDepOfPB}), and Gaussian smearing width $\sigma_g$ (Sec.~\ref{sec:SigmaDepOfPB}).

\subsection{ Dependence of $P_{id}(B)$ on $\lambda $ }\label{sec:LambdaDep}
\begin{figure}[!htb]
\center{\includegraphics[width=0.5\textwidth]{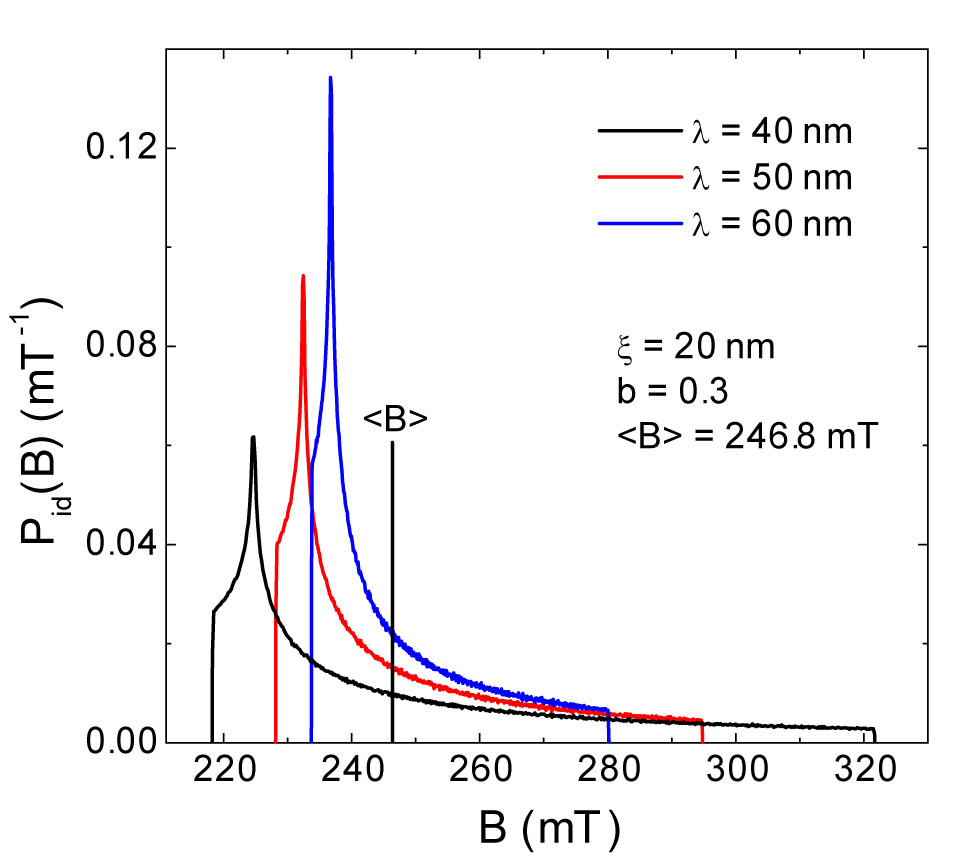}}
\caption{Local magnetic field distribution $P_{id}(B)$  for an ideal hexagonal FLL obtained by the NGL model for different values of $\lambda$, at fixed $\xi$ and applied field $B_{app} \simeq \langle B \rangle$.
The curves are normalized, so that $\int P_{id}(B) \, dB =1$. Note that the shape of $P_{id}(B)$ strongly depends on $\lambda$.}\label{fig:LmdDepShapes}
\end{figure}

In Fig.~\ref{fig:LmdDepShapes} we show examples of the magnetic field distribution $P_{id}(B)$ for different values of the magnetic penetration depth $\lambda$ at constant mean field $\langle B \rangle = 0.3B_{c2} = 246.8$~mT and coherence length $\xi = 20$~nm, as calculated by the NGL model. The region between the minimal and the mean field $\langle B \rangle$ is most important, because the high field tail is usually below the noise level of experimental $\mu$SR spectra and is generally not observed, especially at low fields and for $\kappa \gg 1$. Our calculations show that the differences between the characteristic fields and the mean field $\langle B \rangle$ are proportional to $1/\lambda^2$. This is in full agreement with the results of Sidorenko \textit{et al.} \cite{Sidorenco90} who obtained for applied fields $H_{c1} \ll  H \ll H_{c2}$  and $\kappa \gg 1$ (in this case $\langle B\rangle \simeq \mu_0 H$) in the London approximation the following expressions:
\begin{equation}\label{eqFldMin}
\delta B_{min} = B_{min} - \langle B \rangle = -0.79(\Phi_0/4\pi\lambda^2)\ln2 \, ,
\end{equation}
\begin{equation}\label{eqFldSad}
\delta B_{sad} = B_{sad} - \langle B \rangle = -\frac{2}{3}(\Phi_0/4\pi\lambda^2)\ln2 \, ,
\end{equation}
\begin{equation}\label{eqFldMax}
\delta B_{max} = B_{max} - \langle B \rangle = 2(\Phi_0/4\pi\lambda^2)\ln\frac{a}{2\sqrt{2}K\xi}\, .
\end{equation}
Here, $a$ is the intervortex distance, and $K = K(1/\sqrt{3}) \simeq 1.926$ is the complete elliptic integral of the first kind \cite{Sidorenco90}. Hereafter, for convenience the quantities $\delta B_{min}$, $\delta B_{sad}$, and $\delta B_{max}$ defined above
are denoted as characteristic fields as well. From Fig.~\ref{fig:LmdDepShapes} and the pronounced
dependence of the characteristic fields on $1/\lambda^2$ it is evident that the $\mu$SR time spectra strongly depend on $\lambda$. Therefore, it should be possible to extract reliable values of $\lambda$ from experimental $\mu$SR data.
\begin{figure}[!htb]
\center{\includegraphics[width=0.5\textwidth]{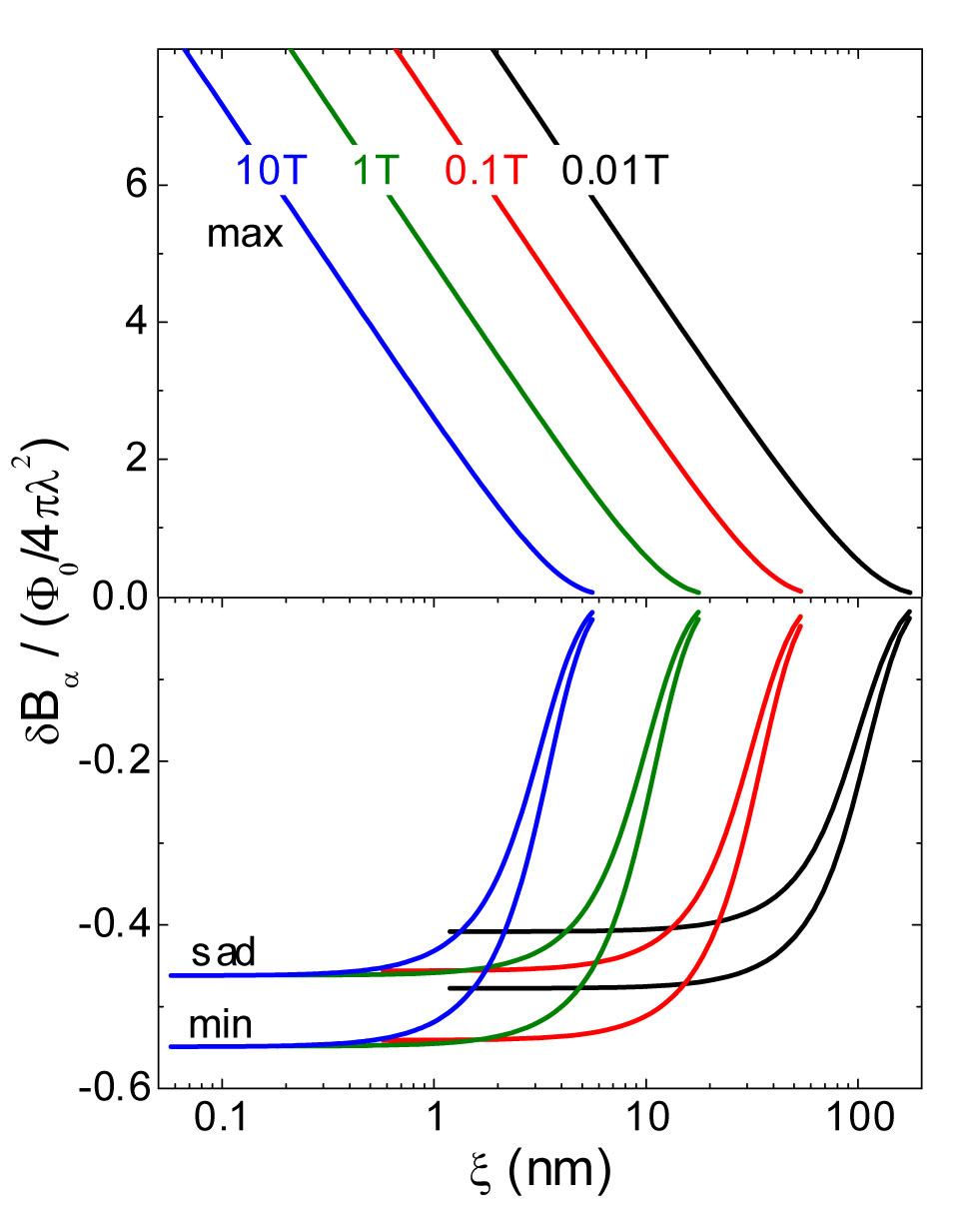}}
\caption{$\xi$ dependence of the characteristic fields $\delta B_\alpha$ ($\alpha$ = min, sad, max) normalized to $\Phi_0/4\pi\lambda^2$ for a set of different applied magnetic fields ($B_{app} \simeq \langle B \rangle $ = 0.01~T, 0.1~T, 1~T, 10~T) as obtained by the LG model. Note that there is a critical value of $\xi$ below which $\delta B_{min}$ and $\delta B_{sad}$ are practically independent of $\xi$. This critical value depends on $\langle B \rangle $.}
\label{fig:XsiDepChFlds}
\end{figure}
\subsection{Dependence of $P_{id}(B)$ on $\xi$}\label{sec:XiDep}
Figure~\ref{fig:XsiDepChFlds} shows the $\xi$ dependence of the characteristic fields $\delta B_\alpha$ ($\alpha$ = min, sad, max) normalized to $\Phi_0/4\pi\lambda^2$ [cf. Eqs.~(\ref{eqFldMin}-\ref{eqFldMax})] for  a set of different mean fields
$\langle B \rangle$, as obtained by the LG model. All the characteristic fields $\delta B_\alpha$ disappear at $\xi \geq (\Phi_0/2\pi \langle B\rangle)^{1/2}$ ($\Phi_0$ is the flux quantum), where superconductivity vanishes. Below a certain value of $\xi$ the characteristic fields $\delta B_{min}$ and $\delta B_{sad}$ are independent of $\xi$, whereas $\delta B_{max}$ still depends on $\xi$. However, in real $\mu$SR experiments $\delta B_{max}$ cannot be determined out of the noise level at low $\langle B \rangle$. Therefore, at these low values of $\xi$ $\mu$SR spectra are practically independent of $\xi$. In order to get a feeling what this means for cuprate superconductors we assume
$\xi \simeq 3$~nm, a typical value of $\xi$ for cuprates below $T_c/2$. In this case the shape of $P_{id}(B)$ is almost independent of $\xi$ for fields $\langle B \rangle \le 0.3$~T, where  $\delta B_{min}$ and $\delta B_{sad}$ saturate (see Fig.~\ref{fig:XsiDepChFlds}). It is thus problematic to find the correct value of $\xi$ at low magnetic fields. At higher fields the shape of $P_{id}(B)$ strongly depends on $\xi$.  Note that in  Fig.~\ref{fig:XsiDepChFlds} the curves corresponding to the smallest field ($\langle B \rangle = 0.01$~T) exhibit slightly smaller saturation values of the characteristic fields $\delta B_{min}$ and $\delta B_{sad}$ than for the higher fields. The reason for this is discussed in Sec.~\ref{sec:FldDepOfPB}.

\begin{figure}[!htb]
\center{\includegraphics[width=0.5\textwidth]{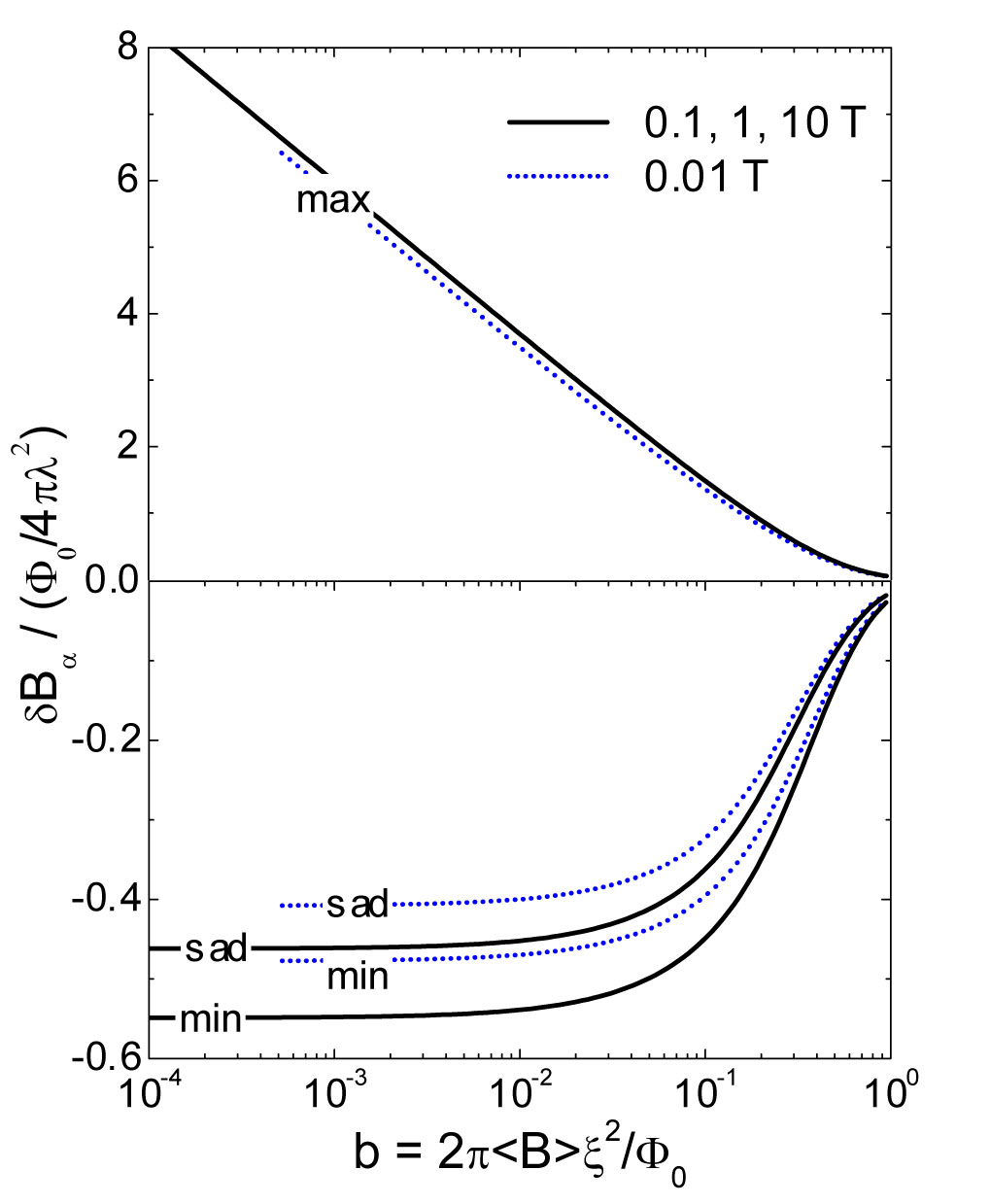}}
\caption{ Characteristic fields $\delta B_\alpha$ ($\alpha$ = min, sad, max) of Fig.~\ref{fig:XsiDepChFlds} plotted as a function of the reduced field $b=\langle B\rangle/B_{c2}(\xi)$ ($B_{c2}(\xi) = \Phi_0/2\pi\xi^2$) at $\langle B \rangle = 0.1$, 1, and 10~T  (black solid line), and at 0.01~T (blue dotted line).  Note that all the curves $\delta B_\alpha(\xi)$ of Fig.~\ref{fig:XsiDepChFlds}, at $\langle B \rangle = 0.1$, 1, 10~T merge to single curves $\delta B_\alpha(b)$.  }
\label{fig:LG_b_depFromXidep}
\end{figure}

As shown by Brandt \cite{Brandt97_NGLmethod, Brandt03} the ideal internal field distribution $P_{id}(B)$ may be expressed by normalized parameters, depending only on $\kappa = \lambda/\xi$ and $b = \langle B\rangle/B_{c2}$.  In a similar way we can plot the curves in Fig.~\ref{fig:XsiDepChFlds} not as a function of $\xi$, but as a function of $b = \langle B\rangle/B_{c2}$, where $B_{c2}(\xi) = \Phi_0/2\pi\xi^2$ (the relation obtained from Ginzburg-Landau theory).  
This plot is shown in Fig.~\ref{fig:LG_b_depFromXidep}.  All the curves of Fig.~\ref{fig:XsiDepChFlds}, except the one for the smallest field $\langle B \rangle = 0.01$~T, fall on the same curve.

\begin{figure}[!htb]
\center{\includegraphics[width=0.5\textwidth]{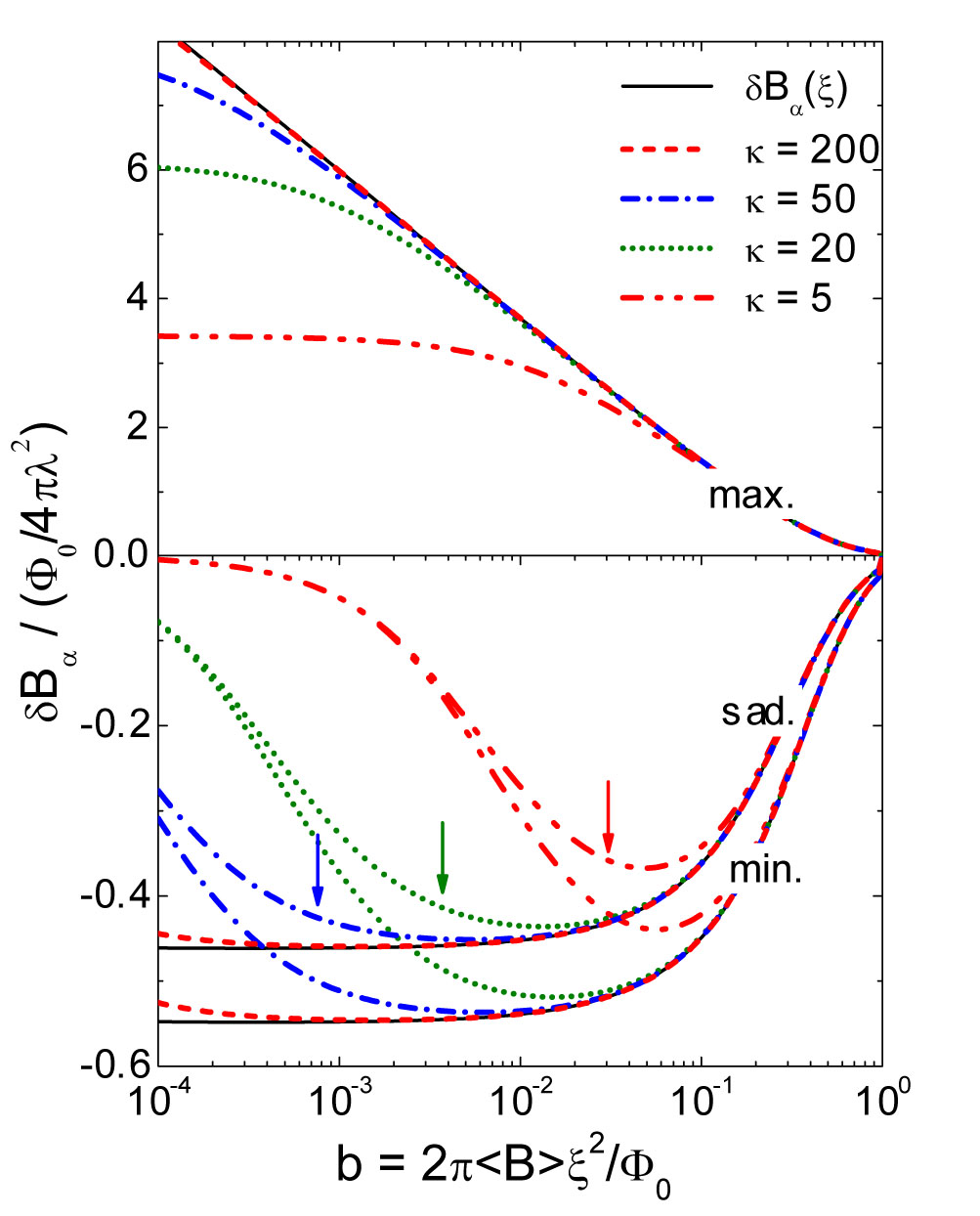}}
\caption{ Characteristic fields $\delta B_\alpha$ ($\alpha$ = min, sad, max) calculated by the LG model as a function of the reduced field $b$ (dashed, dotted and dash-dotted lines) for different values of $\kappa$.  The black solid lines represent the curves $\delta B_\alpha(\xi)$ shown in Fig.~\ref{fig:LG_b_depFromXidep} for $\langle B \rangle = 0.1$~T, 1~T, and 10~T.  The arrows indicate the values of $b_{min}$ at which $\langle B\rangle = B_{c1}$.  }
\label{fig:ML_kappaDep}
\end{figure}
\subsection{Field dependence of $P_{id}(B)$}\label{sec:FldDepOfPB}
Before we discuss the field dependence of the characteristic fields $\delta B_\alpha$ ($\alpha$ = min, sad, max) on various parameters, it is useful to define the minimal value of the reduced field $b_{min} = B_{c1}/B_{c2}\simeq \ln\kappa /2\kappa^2$ which is needed to form a regular FLL. This field corresponds to the limit below which the vortices can be considered as well separated and noninteracting.

Figure~\ref{fig:ML_kappaDep} shows $\delta B_\alpha$ (normalized to $\Phi_0/4\pi\lambda^2$)
as function of the reduced magnetic field $b$ for different values of $\kappa$, as calculated by the LG model. The arrows at $\delta B_{sad}$ correspond to $b_{min}(\kappa)$.  This figure looks very similar to Fig.~\ref{fig:LG_b_depFromXidep} and represents actually its generalization. It shows how $\delta B_\alpha$ depends on all three parameters $\lambda$, $\xi$, $\langle B \rangle$,  and not only $\delta B_\alpha$ as a function of  $\langle B \rangle$. Since Fig.~\ref{fig:ML_kappaDep} demonstrates the  dependence of $\delta B_\alpha$ on all the parameters it is the basis of all further discussions.  At high values of $b$, the characteristic fields $\delta B_{\alpha}(\langle B \rangle)$ and $\delta B_{\alpha}(\xi^{2})$ coincide, but at lower fields they deviate substantially (dependence of $\delta B_{\alpha}$ on a parameter $x$ means that all other parameters, except $x$, are fixed).  The reason for this is obvious.
For $b \rightarrow 0$ at constant $\xi$ or $B_{c2}$ the intervortex distance $a$ increases, and $\langle B \rangle$, $\delta B_{min}(b)$, and $\delta B_{sad}(b)$ tend to zero as well. This is the reason for the smaller saturation values of $\delta B_\alpha$ at the lowest field $\langle B\rangle = 0.01$~T in Fig.~\ref{fig:XsiDepChFlds}. However, in the case of the $\xi$ dependence, when $b \rightarrow 0$ at constant field $\langle B \rangle$, only the vortex core size is reduced, and the intervortex distance $a$ does not change. This has not much influence on the internal magnetic field distribution $P_{id}(B)$ for $\kappa \gg 1$ (see Fig.~\ref{fig:ML_kappaDep}). When  $\kappa$ is reduced, the deviation of $\delta B_{\alpha}(\langle B \rangle)$ from $\delta B_{\alpha}(\langle B \rangle)_{\kappa = \infty}$ starts at higher values of $b$. For small values of $\kappa$ the characteristic fields $\delta B_{\alpha}(\langle B \rangle)$ even do not reach saturated values as in the case of high $\kappa$ and small $b$.
Despite of the similarity of  $\delta B_{\alpha}(\langle B \rangle)$ and
$\delta B_{\alpha}(\xi^2)$, the mean field $\langle B \rangle$ can easily be extracted from the fit (unlike $\xi$), since it defines the oscillation frequency of the $\mu$SR time spectrum.
\begin{figure}[!htb]
\center{\includegraphics[width=0.5\textwidth]{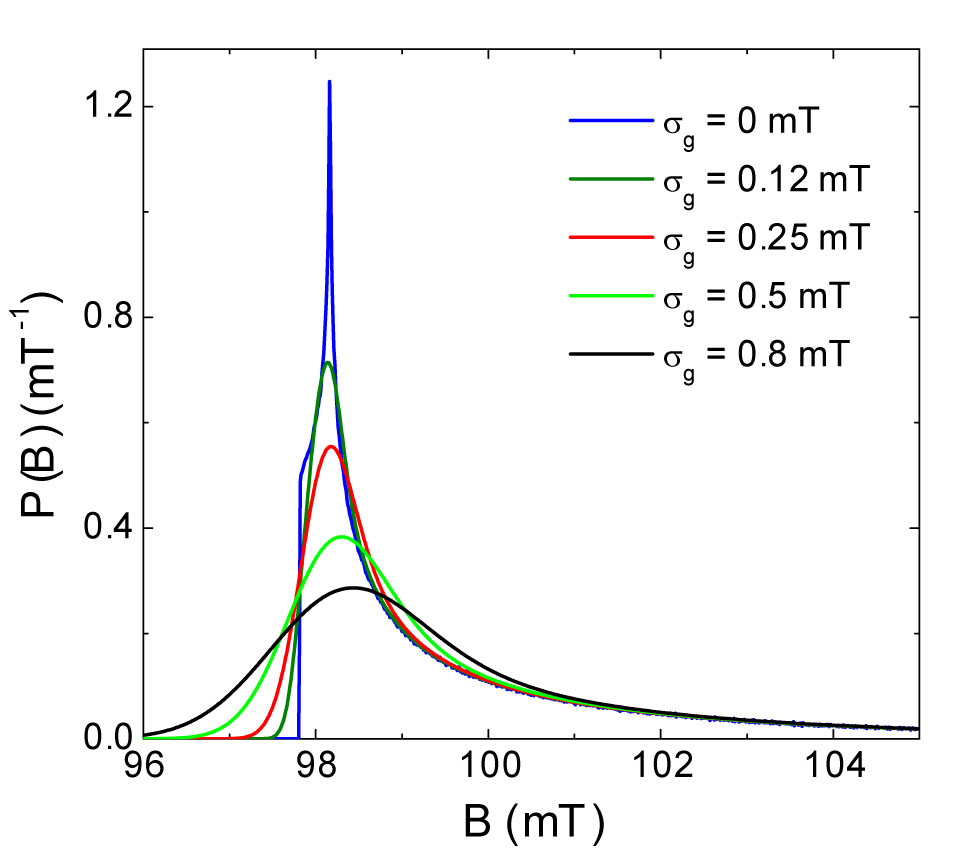}}
\caption{ Change of the local magnetic field distribution $P_{id}(B)$ for an ideal FLL after convolution with a Gaussian distribution of various widths $\sigma_g$. The following parameters were used to generate $P_{id}(B)$
with the LG model: $\lambda = 200$~nm, $\xi = 4$~nm,
and $\langle B \rangle = 100$~mT. }
\label{fig:SigDepShape}
\end{figure}
\subsection{ Dependence of $P(B)$ on $\sigma_g$ }\label{sec:SigmaDepOfPB}
In reality the internal magnetic field distribution in the mixed state of a type-II superconductor is influenced by several factors, which generally lead to an additional broadening of $P_{id}(B)$: (i) The FLL is never ideal, but disordered by random pinning effects of the vortex cores. (ii) For powder samples of anisotropic superconductors - such as the layered cuprate superconductors - the grains usually have random shapes, and therefore have  anisotropic superconducting properties and demagnetization effects play a role \cite{Weber93}. (iii) The sample may contain magnetic nuclear moments or paramagnetic impurities.   Vortex disorder and nuclear broadening can be taken into account by convoluting the ideal internal field distribution $P_{id}(B)$  with a Gaussian distribution of width (see Eq.~\ref{EqConvolution}) \cite{BrandtJLTPhys88and77, Riesman95, Reotier04}:
\begin{equation}\label{eq:SigmaG}
\sigma_g = \sqrt{\sigma_{VD}^2 + \sigma_{N}^2},
\end{equation}
where $\sigma_{VD}$ and $\sigma_N$ are the contributions to the Gaussian broadening of $P_{id}(B)$ due to vortex disorder and nuclear depolarization, respectively \footnote{For powder samples with a Gaussian distribution of the first moments $\langle B\rangle$ due to Gaussian distribution of demagnetization factors $N$ one may add an additional term $\sigma^2_{\langle B\rangle}$ to Eq.~(15).}.
For $\kappa \gg 1$ the standard deviation of the vortex core positions from the ideal positions in
the FLL $\left<s^2\right>^{1/2}$ is related to  $\sigma_{VD}$ by the following equation \cite{Riesman95}:
%
%
%
\begin{equation}
\sigma_{VD} \propto \lambda^{-2}\cdot (1-b) \cdot \frac{\left<s^2\right>^{1/2}}{a}\,
\label{eq:SigmaG2}
\end{equation}
Here $b = \langle B \rangle/B_{c2}$, and $a$ is the intervortex distance. 

Figure \ref{fig:SigDepShape} shows examples of  $P(B)$ for $\lambda = 200$~nm, $\xi = 4$~nm, $\langle B \rangle = 0.1$~T, and for various values of $\sigma_g$, calculated by
means of the LG model.
It is obvious that with increasing degree of disorder the Van Hove  singularities in the ideal internal field distribution $P_{id}(B)$ are smeared out. Note that the low-field part of $P_{id}(B)$ is mainly truncated by the Gaussian smearing, whereas the high-field tail is nearly not affected.

\section{Comparison of different models} \label{sec:ComparisonOfDifferentModels}

\begin{figure}[tb]
\center{\includegraphics[width=0.5\textwidth]{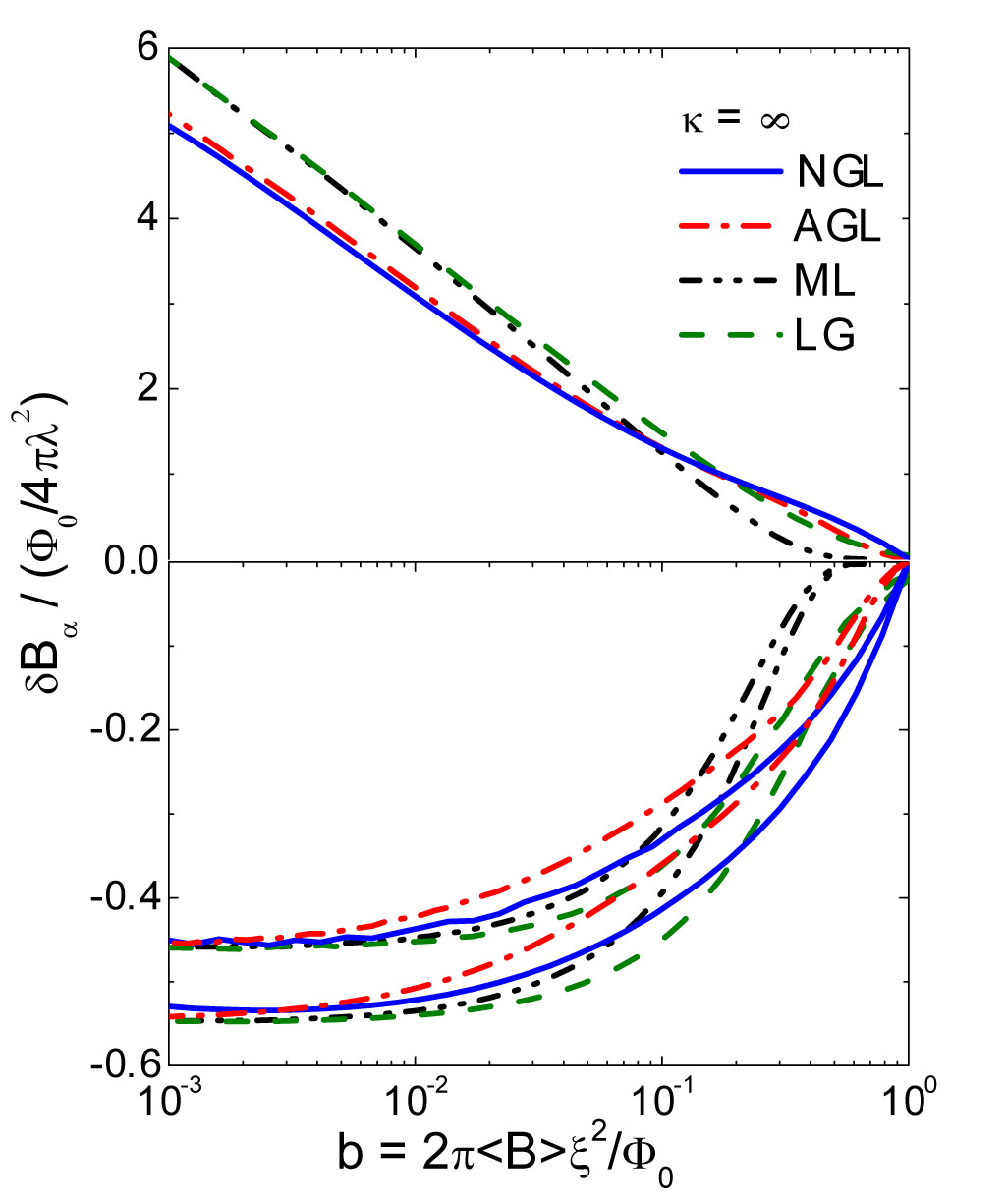}}
\caption{ Characteristic fields $\delta B_\alpha$ ($\alpha$ = min, sad, max) as a function of the reduced field $b = 2\pi \langle B \rangle \xi^2/\Phi_0$ for $\kappa = \infty$, as calculated by the LG, ML, AGL, and NGL model. }
\label{fig:ModelComp}
\end{figure}

In this section the different models (LG, ML, AGL, and NGL) discussed in this work are compared. For this purpose the NGL model, is used as a reference model to describe the mixed state of a type-II superconductor. In Sec.~II we showed that the characteristic fields $\delta B_\alpha$ for $\kappa \gg 1$
may be represented by single curves (see Fig.~\ref{fig:ML_kappaDep}).
Figure~\ref{fig:ModelComp} shows the characteristic fields $\delta B_\alpha$ as a function of the reduced magnetic field $b=\langle B\rangle/B_{c2}$ in the limit of $\kappa  \rightarrow \infty$ as calculated by the LG, ML, AGL, and NGL model. For small $b$ values $\delta B_{min}(b) $ and $\delta B_{sad}(b) $ coincide for all models.
Deviations of the AGL and LG model from the NGL model appear above
$b \approx 0.01$. Although the AGL and LG models may well fit the $\mu$SR spectra simulated by the NGL model, the fitted values of $\xi$ may  deviate substantially from the {\em real values} for reduced magnetic fields $b \gtrsim 10^{-2} $.
This systematic deviation increases with increasing magnetic field.
For the LG model, in contrast to the AGL model, the systematic errors even change sign with increasing magnetic field. One should note, that for $\kappa > 5$ there is no advantage of using the AGL model instead of the LG model. Of all the models the  ML model approximates best the NGL model up to about $b \simeq 0.1$, in agreement with previous results of Brandt \cite{Brandt88PRB}.  However, at higher fields this model substantially deviates from the NGL model. The ML model was often used to analyze experimental data for $b > 0.1$ \cite{Sonier00RMP, Laulajainen06}. We found that $\mu$SR spectra simulated by the NGL model in the range $b = 0.1 - 1$ may well be fitted by the ML model. But for $b > 0.1$ the values of $\xi$ extracted from the simulated $\mu$SR spectra are artificially reduced compared to the real values of $\xi$ (see Fig.~\ref{fig:ModelComp}).
For smaller values of $\kappa$, the characteristic fields $\delta B_\alpha$ for the ML, AGL, and NGL models behave similarly to those  of the LG model (see Fig.~\ref{fig:ML_kappaDep}). Namely, for high reduced fields $b$ all the curves with different values of $\kappa$ coincide. The smaller $\kappa$ is, the higher is the reduced field $b$ when they start to deviate from the curves shown in Fig.~\ref{fig:ModelComp}.
The only exception is the AGL model for which for $\kappa \le 5$ the curves become closer to the NGL curves. Here we should mention that our results obtained with the NGL model are in full agreement with the calculations of Brandt \cite{Brandt03}.
\section{Simulation and fitting of $\mu$SR spectra}
%

In order to check the conclusions we made in the previous sections "experimentally",
$\mu$SR time spectra with known parameters ($\langle B\rangle, \lambda, \xi, \sigma_g$) were simulated using the LG and the NGL models. For the simulation of the $\mu$SR experiment a transverse-field (TF) configuration with two positron detectors D1 and D2 located on opposite sides of the sample was used.
The number of positrons detected by the detector D1 at time $t_i = i \cdot \Delta t$ is $N_1(t_i)\Delta t$   ($i = 1,2,...,8000$; $\Delta t$ was chosen to be 1.25~ns, corresponding to a typical time-resolution for the GPS spectrometer at the Paul Scherrer Institute, Switzerland).
This positron count number obeys Poisson statistics, and the standard deviation is simply given by $\sqrt{N_1(t_i)\Delta t}$. In the ideal case of no noise, the detector D1 would detect the signal $n_1(t_i) = n_0 e^{-t_i/\tau_\mu}[1 + P(t_i)] $,
where $n_0$ is a constant depending on the number of muons detected (statistics) and the time interval $\Delta t$, $\tau_\mu = 2.197019(21)$~$\mu$s is the muon lifetime, and $P(t_i)$ is the noiseless $\mu$SR  time signal [see Eq. (\ref{EqTdomainSpec})] to which noise has to be added. The signal monitored by detector D1 can be simulated by the equation   $N_1(t_i) = n_1(t_i) + \sqrt{n_1(t_i)}g_i $, where $g_i$ is a random number generator obeying Gaussian statistics with standard deviation and variance equal to 1. A similar signal but with opposite phase is registered by detector D2. In analogy to real $\mu$SR experiments one can calculate  the asymmetry \cite{ShenkBook, Sonier00RMP} $A(t_i) = [N_1(t_i)-N_2(t_i)]/[N_1(t_i)+N_2(t_i)]$, yielding $P(t_i)$ with "experimental noise". The $\mu$SR time spectra were simulated according to the procedure described above with total statistics of 20 million events, a value typically used in real experiments.

The simulated $\mu$SR time spectra were then analyzed as follows:\\
(1) The $\mu$SR spectra simulated by the NGL model were analyzed by the second-moment (SM) method.\\
(2) The $\mu$SR spectra simulated by the LG model were analyzed by a fitting procedure
based on the LG model.

According to the discussions made in the previous sections the following important points emerge:
~\\
(1) How reliable is the second moment obtained by a multi-Gaussian fit of the $\mu$SR spectra (see Eq. (\ref{eq:multi-gauss})) and the value of the penetration depth $\lambda$ extracted from the second moment?
~\\
(2) Is there a correlation between $\sigma_g$ and $1/\lambda^2$, since both of them influence the second moment of the $\mu$SR spectrum?
~\\
(3) Is it possible  to extract reliable values of $\xi$ from $\mu$SR spectra at low magnetic fields $b \simeq 10^{-3}$?
~\\
(4) Is there a correlation between $\lambda$ and $\xi$ at high fields? Since for $b \approx 0.1 - 0.9$  both parameters strongly influence the characteristic fields $\delta B_\alpha$.
\subsection{Test of the second-moment method}\label{sec:TestOfSMmethod}
In this section the second-moment (SM) method is tested by analyzing $\mu$SR time spectra simulated with the NGL model with well defined parameters ($\langle B \rangle$, $\lambda$, $\xi$, $\sigma_g$). The SM method is theoretically well described in the literature \cite{Brandt88PRB, Sidorenco90, Brandt03} and was extensively used to extract the magnetic penetration depth of extreme type-II superconductors from $\mu$SR spectra \cite{Aeppli87, Harshman87, Uemura88,  Pumpin90, Zimmermann95, Aegerter97, Lee99}.
In the framework of this method the $\mu$SR time spectra are usually fitted to a sum of $N$ Gaussian components \cite{Weber93, Khasanov06TwoGap}:
\begin{equation}
P(t) = \sum_{i=1}^{N}A_i\exp(-\sigma_i^2t^2/2)\cos(\gamma_\mu B_it + \phi).
\label{eq:multi-gauss}
\end{equation}
Here $\phi$ is initial phase of the muon beam, and $A_i$, $\sigma_i$, and $B_i$ are the asymmetry, the relaxation rate, and the first moment of the $i-$th Gaussian component, respectively. From the fit parameters $A_i$, $\sigma_i$, and $B_i$ one readily obtains the first and the second moment of $P(B)$ from Eqs. (\ref{eq:Baver}) and~(\ref{eq:dB}), respectively. Using Eq.~(\ref{eq:sec-mom_general}) one finds the penetration depth $\lambda$. Here the serious question arises: How reliable                                                                                                             is the value of $\lambda$ obtained by the SM method using a multi-Gaussian fit?
In order to investigate this question $\mu$SR time spectra were simulated by the NGL model for an extreme type-II superconductor (such as the cuprate superconductors) with a Ginzburg-Landau parameter $\kappa = 50 \gg 1$. The temperature dependence of $\lambda$ was assumed to follow the relation (two-fluid model):  $\lambda(T)^{-2}/\lambda(0)^{-2} = [1-(T/T_c)^4]$ with $T_c = 22.5$~K and $\lambda(0) = 200$~nm (zero-temperature penetration depth). This approximately corresponds to the temperature dependence of $\lambda$ in the strong-coupling BCS case \cite{TinkhamBook96}.  In the first step we assume that there is no vortex disorder ($\sigma_{VD} = 0$) and no nuclear depolarization ($\sigma_N = 0$) present [cf. Eq.~(\ref{eq:SigmaG})]. The simulations were performed for three different magnetic fields: $\langle B \rangle$ = 0.05, 0.5, and 5~T \footnote{In a real $\mu$SR experiment with a time binning $\Delta t = 1.25$~ns the maximal possible field for a measurement is 2.95~T (this corresponds to two binnings per precession period). At higher fields the asymmetry of the signal drops. In our simulations, unlike in experiments, we do not integrate the positron counts within the time interval $\Delta t$, but simulate the detector counts at $t_i = i\cdot \Delta t$. For this reason it is possible to simulate and fit $\mu$SR data at higher  fields due to the stroboscopic effect. In this case the absolute value of the real field is the fitted field plus $N\cdot 2.95$~T, where $N$ is a positive integer number.}. 
This corresponds to the reduced fields $b$ = 0.0024, 0.024, and 0.24, an extremely small, an intermediate, and a high magnetic field, respectively (see Fig.~\ref{fig:ML_kappaDep}).  Since $B_{c2}$ is decreasing with increasing temperature (1/$\xi^2 \propto B_{c2}$; $1/\xi^2 \propto 1/\lambda^2$ at constant $\kappa$), the analysis of the $\mu$SR spectra for $B = 5$~T makes sense only up to 21~K where $B_{c2}(21K) \approx \langle B\rangle$ and superconductivity disappears.  Noisy $\mu$SR time spectra were simulated with the parameters $\lambda$, $\xi$, and $\langle B \rangle$ as described above. For the technical parameters the following typical values were used: statistics $20\cdot 10^6$, asymmetry $A = 0.2$, and phase $\phi = 0$.

\begin{figure}[tb]
\center{\includegraphics[width=0.5\textwidth]{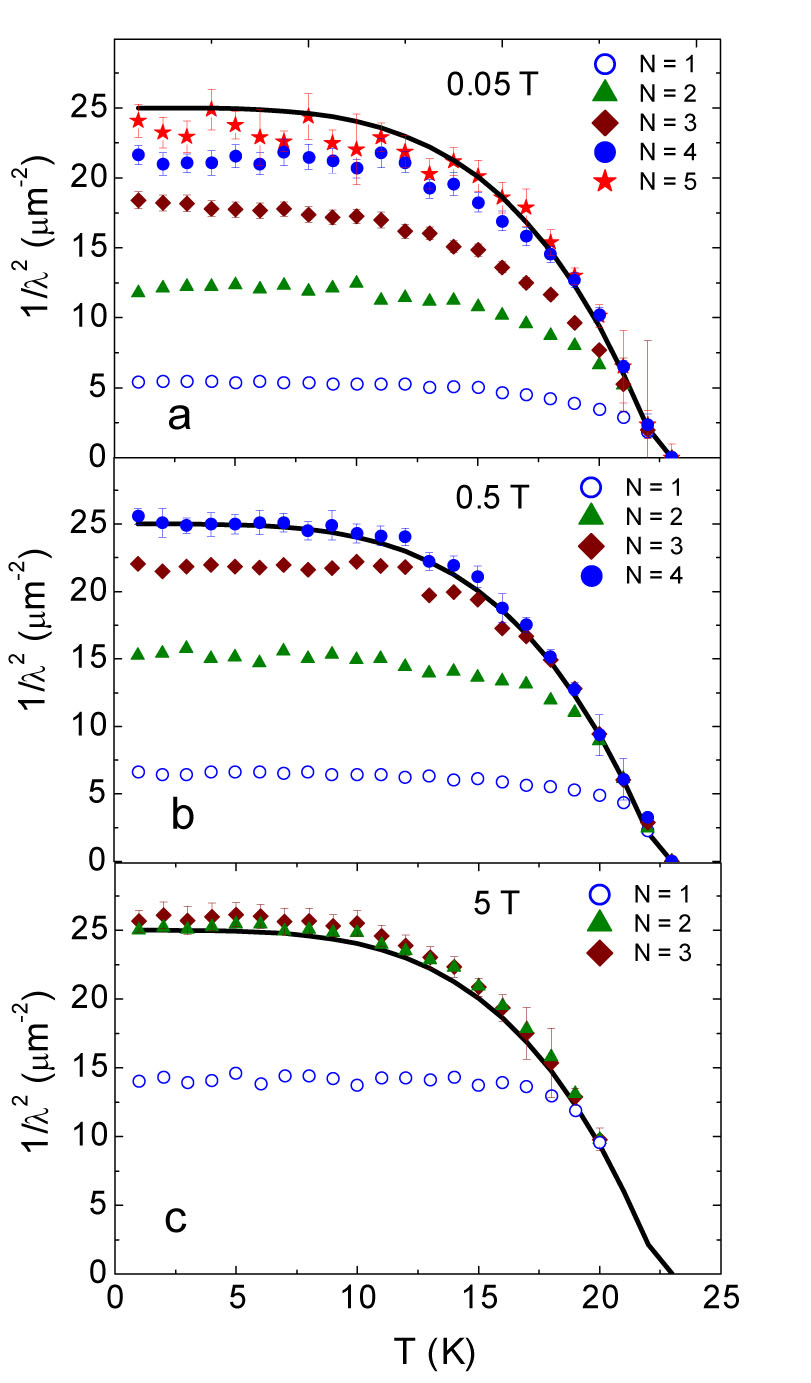}}
\caption{  Fit results for $\lambda^{-2}$ obtained by the second moment method. The noisy spectra for the three different fields of 0.05, 0.5, and 5~T  were simulated by the NGL method for an ideal FLL as described in the text, and then analyzed by a multi-Gaussian function with different number of Gaussians ($N = 1,2,3,4,5$) as defined in Eq.~(\ref{eq:multi-gauss}). The black solid lines correspond to the real values of $\lambda^{-2}$ used for the simulation.   }\label{fig:SMzeroSigma1}
\end{figure}

\begin{figure}[tb]
\center{\includegraphics[width=0.5\textwidth]{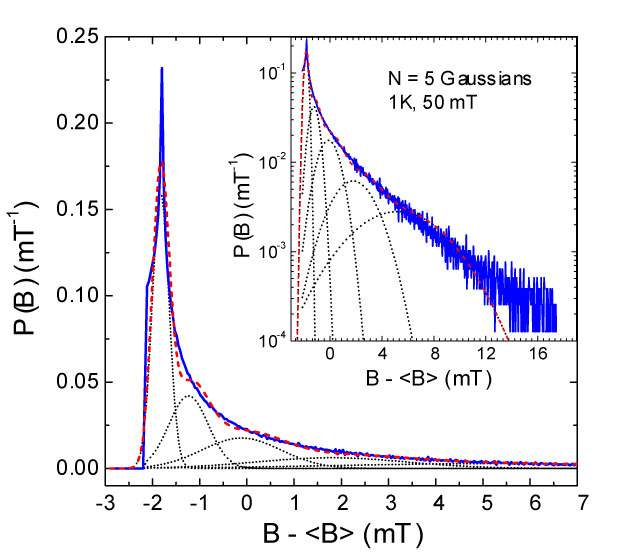}}
\caption{   Comparison of the ideal field distribution $P_{id}(B)$ simulated with the NGL model at 0.5~T and 1~K (blue line) with $P(B)$ obtained by a Gaussian fit with $N = 5$ (red dashed line). The five individual Gaussian components  used for the fit are also shown (black dotted lines). The inset shows the same plot but on a semi-logarithmic scale.  }\label{fig:Fit5GssIdeal}
\end{figure}

For the analysis of the simulated $\mu$SR spectra, $P(t)$ in Eq.~(\ref{eq:multi-gauss}) was approximated by a sum of $N = 1, 2, 3, 4, 5$ Gaussians in order to check the reliability of the result obtained by a multi-Gaussian fit. The number of Gaussians $N$ determines the quality of the fit. $N$ should be increased from 1 until $\chi^2$ (normalized to the degrees of freedom) is close to 1 within statistical scattering. The total asymmetry $A = \sum_{i=1}^{N} A_i$ and the phase $\phi$ of $P(t)$ in  Eq.~(\ref{eq:multi-gauss}) were assumed to be known and were fixed in the fitting procedure. According to our experience, in order to reduce the scattering of the fitted values of the second moment, one should fix the asymmetries $A_i$ of the individual Gaussians to their average values obtained by a fit with all parameters free.  From the first and the second moments of the individual Gaussians one can calculate the second moment $\langle \Delta B^2 \rangle$ of the $\mu$SR spectrum using Eq.~(\ref{eq:dB}), that corresponds to the second moment of $P_{id}(B)$ of an ideal FLL.  The magnetic penetration depth $\lambda$ is readily obtained from $\langle \Delta B^2\rangle$ with Eq.~(\ref{eq:sec-mom_general}).  
The result of the analysis of the  simulated $\mu$SR time spectra for an ideal FLL is shown in Fig.~\ref{fig:SMzeroSigma1}. For the 0.05~T data the finite value of $\kappa = 50$ was taken in account in the coefficient $C$ in Eq.~(\ref{eq:sec-mom_general}) \cite{Brandt03}.   One can see that the smaller the field is the more Gaussians $N$ are needed to describe the spectra. Whereas for 5~T $N = 2, 3$ Gaussians are sufficient to reproduce the spectra, $N=4$ and $N=5$ are required for 0.5 and 0.05~T, respectively. Note that the scattering of the data points increases with increasing number of Gaussians $N$. Although at 0.05~T the fitted values of $1/\lambda^2$ deviate systematically from the real values by a few percent, the qualitative behavior of $1/\lambda^2(T)$ is the same. As will be shown below by adding a Gaussian smearing $\sigma_g$ to the $\mu$SR spectra, the scattering is reduced, and a smaller number of Gaussians $N$ are needed to describe the spectra.  Fig.~\ref{fig:Fit5GssIdeal} demonstrates how the local magnetic field distribution $P_{id}(B)$ for an ideal FLL can be approximated by $N = 5$ Gaussians. Although not all the details of $P_{id}(B)$ are reproduced, the overall agreement is good, in particular the second moment.

In order to test the second moment method under more realistic conditions one should add a Gaussian smearing $\sigma_g$ to the $\mu$SR spectra [cf. Eq.~(\ref{EqConvolution})]. According to Eq.~(\ref{eq:SigmaG}), we assume for the further discussions that $\sigma_g$ is composed of two components: $\sigma_g = (\sigma_{VD}^2 + \sigma_N^2)^{1/2}$, where $\sigma_{VD}$ denotes the temperature dependent smearing due to vortex disorder, and $\sigma_N$ is the temperature independent smearing due to nuclear depolarization  [cf. Eq.~(\ref{eq:SigmaG2})].
For a constant vortex disorder $\left<s^2\right>^{1/2}/a = const.$ (rigid vortex lattice) and reduced field $b$, the relation $\sigma_{VD} \propto 1/\lambda^2$ holds. 
As is obvious from Eq.~(\ref{eq:SigmaG2}), for $ \langle B\rangle = 5$~T the term $(1-b) = 0.76$  substantially deviates from unity and has to be taken into account in the simulation of $\mu$SR data.

For the simulations of the smeared $\mu$SR spectra the following values for single crystal La$_{1.83}$Sr$_{0.17}$CuO$_{4-\delta}$ were used: $\sigma_{VD} = \beta/\lambda^2$, $\beta = 2.585\cdot 10^{4}$~(mT$\cdot$nm$^2$) and $\sigma_{N} = \sigma_{Cu} = 0.27$~mT \cite{me}. Noisy $\mu$SR time spectra were simulated with the parameters $\lambda$, $\xi$, $\langle B \rangle$, and $\sigma_g$ as described above. For the technical parameters the following typical values were used: statistics $20\cdot 10^6$, asymmetry $A = 0.2$, and phase $\phi = 0$.

\begin{figure}[tb]
\center{\includegraphics[width=0.5\textwidth]{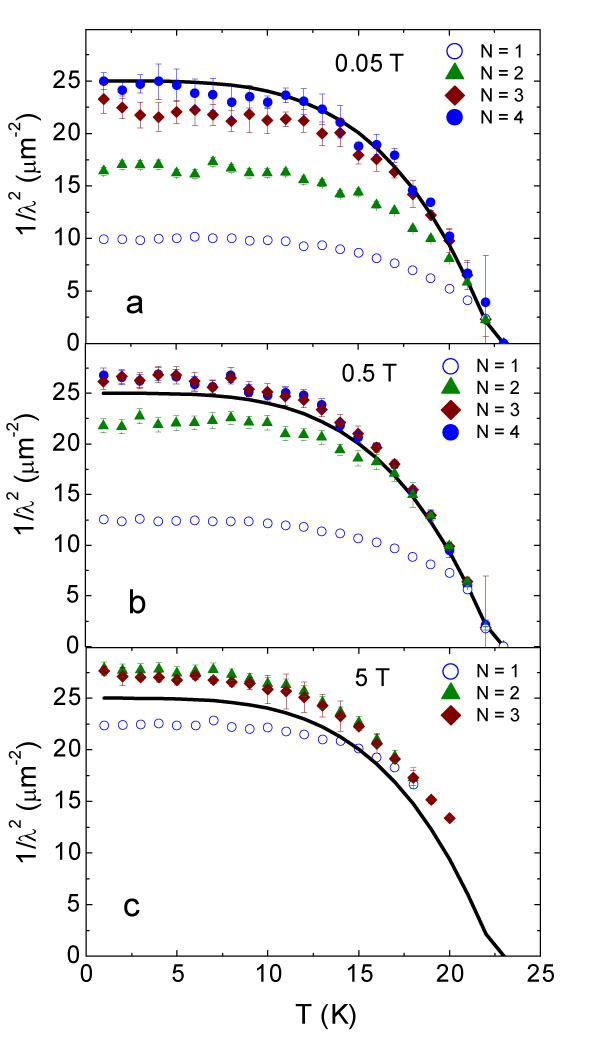}}
\caption{ Fit results for $\lambda^{-2}$ obtained by the second moment method. The noisy spectra for the three different fields of 0.05, 0.5, and 5~T  were simulated by the NGL method including Gaussian smearing $\sigma_g$ as described in the text, and then analyzed by a multi-Gaussian function with different number of Gaussians ($N = 1,2,3,4$), as defined in Eq.~(\ref{eq:multi-gauss}). The black solid lines correspond to the real values of $\lambda^{-2}$ used for the simulation.   }\label{fig:multiGssTdep}
\end{figure}

\begin{figure}[tb]
\center{\includegraphics[width=0.5\textwidth]{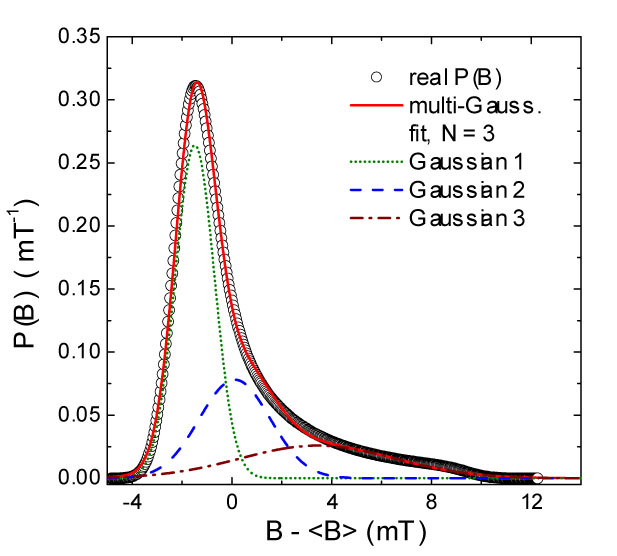}}
\caption{ Comparison of the real field distribution $P(B)$ (empty circles) simulated with the NGL model for parameters $\lambda = 200$~nm, $\xi = 4$~nm, $\langle B\rangle = 0.5$~T, $\sigma_g = 0.7$~mT (data point at $T=1$~K for $N = 3$ in Fig.~\ref{fig:multiGssTdep}b) with $P(B)$ obtained by a Gaussian fit with $N = 3$ (red line). The dotted, dashed, and dash-dotted lines represent individual Gaussian components  used for the fit. It is obvious that a  multi-Gaussian fit may well describe the real $P(B)$.  }\label{fig:multiGss_0.5T_0K}
\end{figure}

The total second moment $\langle \Delta B^2 \rangle_t$ of a $\mu$SR spectrum with Gaussian smearing $\sigma_g$ obtained by multi-Gaussian fit contains three components:
\begin{equation}\label{eq:sec-mom-LDN}
\langle \Delta B^2 \rangle_t = \langle \Delta B^2 \rangle  +  \sigma_{VD}^2  +  \sigma_N^2,
\end{equation}
where $\langle \Delta B^2 \rangle$, $\sigma_{VD}^2$, and $\sigma_N^2$ are the second moments due to the  internal field variation in the ideal FLL, the vortex disorder, and the nuclear depolarization, respectively. In order to obtain  $\lambda$ from the total second moment measured in real experiments one determines $\sigma_N$ above $T_c$, and assumes that $\sigma_{VD}^2 \ll \langle \Delta B^2 \rangle$  in Eq.~(\ref{eq:sec-mom-LDN}), i.e. $\langle \Delta B^2 \rangle_t \simeq \langle \Delta B^2 \rangle  +  \sigma_N^2$ \cite{Khasanov06LiPd}. From the first and the second moments of the individual Gaussians one can calculate the total second moment $\langle \Delta B^2 \rangle_t$ of the $\mu$SR spectrum using Eq.~(\ref{eq:dB}) [Note that in Eq.~(\ref{eq:dB}) $\langle \Delta B^2 \rangle$ has to be replaced by $\langle \Delta B^2 \rangle_t$ for the case $\sigma_g \neq 0$]. By means of  Eqs.~(\ref{eq:sec-mom-LDN}) and (\ref{eq:sec-mom_general}) the magnetic penetration depth $\lambda$ then is readily obtained.
Figure~\ref{fig:multiGssTdep} shows the results for the penetration depth obtained by the second moment method with $N = 1$, 2, 3, 4 Gaussians. Note that a single Gaussian does not give reliable results in agreement with earlier findings \cite{Weber93}.  However, with increasing number of Gaussians $N$ the quality of the fits substantially improves. In order to fit the simulated data at 0.05~T at least 3 or better 4 Gaussians are required. For $N = 3$  there is a systematic deviation of about 10\% of $\lambda^{-2}$  from the real value (or 5\% for $\lambda$), whereas for $N = 4$ the values of $\lambda^{-2}$ are scattered within a few percent around the real ones. For the data simulated at 0.5~T even $N=3$ Gaussians are sufficient to describe the local magnetic field distribution $P(B)$, and the values of $\lambda^{-2}$ are systematically shifted only within a few percent. Fig.~\ref{fig:multiGss_0.5T_0K} shows an example of a real internal field distribution $P(B)$ ($\langle B\rangle = 0.5$~T, $T = 1$~K of Fig.~\ref{fig:multiGssTdep}) and the reconstructed $P(B)$ obtained from the analysis of the simulated $\mu$SR spectrum (2$\times10^{7}$ statistics) using a Gaussian fit with $N = 3$. It is obvious that three Gaussians describe well the shape of the real $P(B)$. The largest systematic error in $\lambda^{-2}$ obtained by a multi-Gaussian fit is observed at 5~T ($b = 0.24$). At such a high field ($\langle B\rangle \simeq B_{c2}/4$) the variation of the internal field is relatively small (see Fig.~\ref{fig:ModelComp}), and the Gaussian smearing $\sigma_{VD}$ [cf. Eq.~(\ref{eq:sec-mom-LDN})] due to vortex disorder becomes essential. The second moment of this Gaussian smearing cannot be neglected and considerably contributes to the total second moment $\langle \Delta B^2 \rangle_t$ of $P(B)$. This leads to systematically higher values of $\lambda^{-2}$ obtained by multi-Gaussian fits at high magnetic fields. Note, however, there are two reasons  why the contribution to the second moment due to vortex disorder is reduced with increasing magnetic field, and consequently the systematic error in $\lambda^{-2}$: (1) At high fields $\sigma_{VD} \propto (1-b)$ [cf. Eq.~(\ref{eq:SigmaG2})], which was taken into account in the simulations of the $\mu$SR spectra. (2) Vortex disorder, $\left<s^2\right>^{1/2}/a$, is expected to decrease with increasing magnetic field, because of the strong repulsive interaction between the vortices at high fields (see e.g. Ref.~\onlinecite{Brandt03}).

\begin{figure}[!htb]
\clearpage
\center{\includegraphics[width=0.5\textwidth]{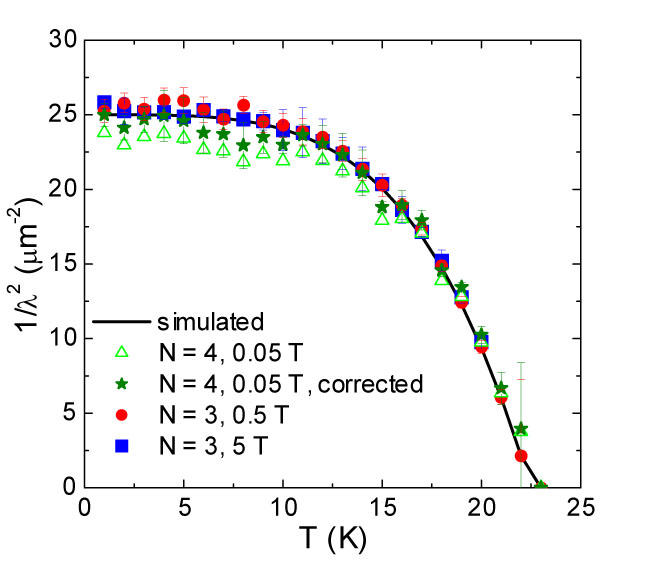}}
\caption{ Temperature dependence of $1/\lambda^2$ determined from $\langle \Delta B^2\rangle$, assuming that $\sigma_{VD}$ is known as described in the text. Triangles: $N = 4$, 0.05~T; stars: $N=4$, 0.05~T, corrected for finite $\kappa = 50$ (see text for explanation); circles: $N=3$, 0.5~T; squares: $N = 3$, 5~T.   }\label{fig:second-mom-corrected}
\end{figure}

By means of a multi-Gaussian fit it is not possible to separate $\langle \Delta B^2 \rangle$ and $\sigma_{VD}$ from the measured total second moment $\langle \Delta B^2 \rangle_t$ [cf. Eq.~(\ref{eq:sec-mom-LDN})]. Assuming that $\sigma_{VD} = 0$ yields a lower limit for $\lambda$ (upper limit for $1/\lambda^2$), as clearly demonstrated in Figs.~\ref{fig:multiGssTdep}(b) and \ref{fig:multiGssTdep}(c) where the values of $1/\lambda^2$ are systematically too large. It is interesting to investigate what are the values of $1/\lambda^2$ after correction with the real value of $\sigma_{VD}$. For this purpose we write Eq.~(\ref{eq:SigmaG2}) in the form $\sigma_{VD} = \beta \lambda^{-2}$ and with the help of Eqs.~(\ref{eq:sec-mom_general}) and (\ref{eq:sec-mom-LDN}) we obtain:
\begin{equation}\label{eq:sec-mom-general-FLLcorrected}
\lambda^{-2} = [C/(1+\epsilon)]^{1/2}[\langle \Delta B^2 \rangle_t - \sigma_N^2]^{1/2},
\end{equation}
where $\epsilon = C\beta^2$ describes the correction due to vortex disorder. The values of $1/\lambda^2$ plotted in Fig.~\ref{fig:multiGssTdep} were obtained with $\epsilon = 0$ (no vortex disorder correction). Fig.~\ref{fig:second-mom-corrected} shows some of the results of Fig.~\ref{fig:multiGssTdep} after correcting the values of $1/\lambda^2$ with the values of $\epsilon_{T=0} = 0.050$, 0.073, and 0.155 for 0.05~T, 0.5~T, and 5~T, respectively  ($\epsilon$ is temperature dependent, since $C(b(T),\kappa)$ is temperature dependent). The corrected values of $1/\lambda^2$ are in good agreement with the real values (solid line in Fig.~\ref{fig:second-mom-corrected}), except for the data at 0.05~T where a systematic deviation of about 5-10\% is observed. One of the reasons for this deviation  is that $\kappa = 50$ used in the simulations is not infinite. This implies that at 0.05~T ($b\simeq 0.0024$) the parameter $C^{-1/2}$ in Eqs.~(\ref{eq:sec-mom_general}) and (\ref{eq:sec-mom-general-FLLcorrected}) is about 5\% smaller \cite{Brandt03}.  The stars in Fig.~\ref{fig:second-mom-corrected} represent the corrected values of $1/\lambda^2$ at 0.05~T, which are only about 3\% systematically lower than the real values.

%
%
%
\subsection{Test of the London model with Gaussian cut off (LG)}\label{sec:TestOfLGm}

\begin{figure}[!htb]
\clearpage
\center{\includegraphics[width=0.45\textwidth]{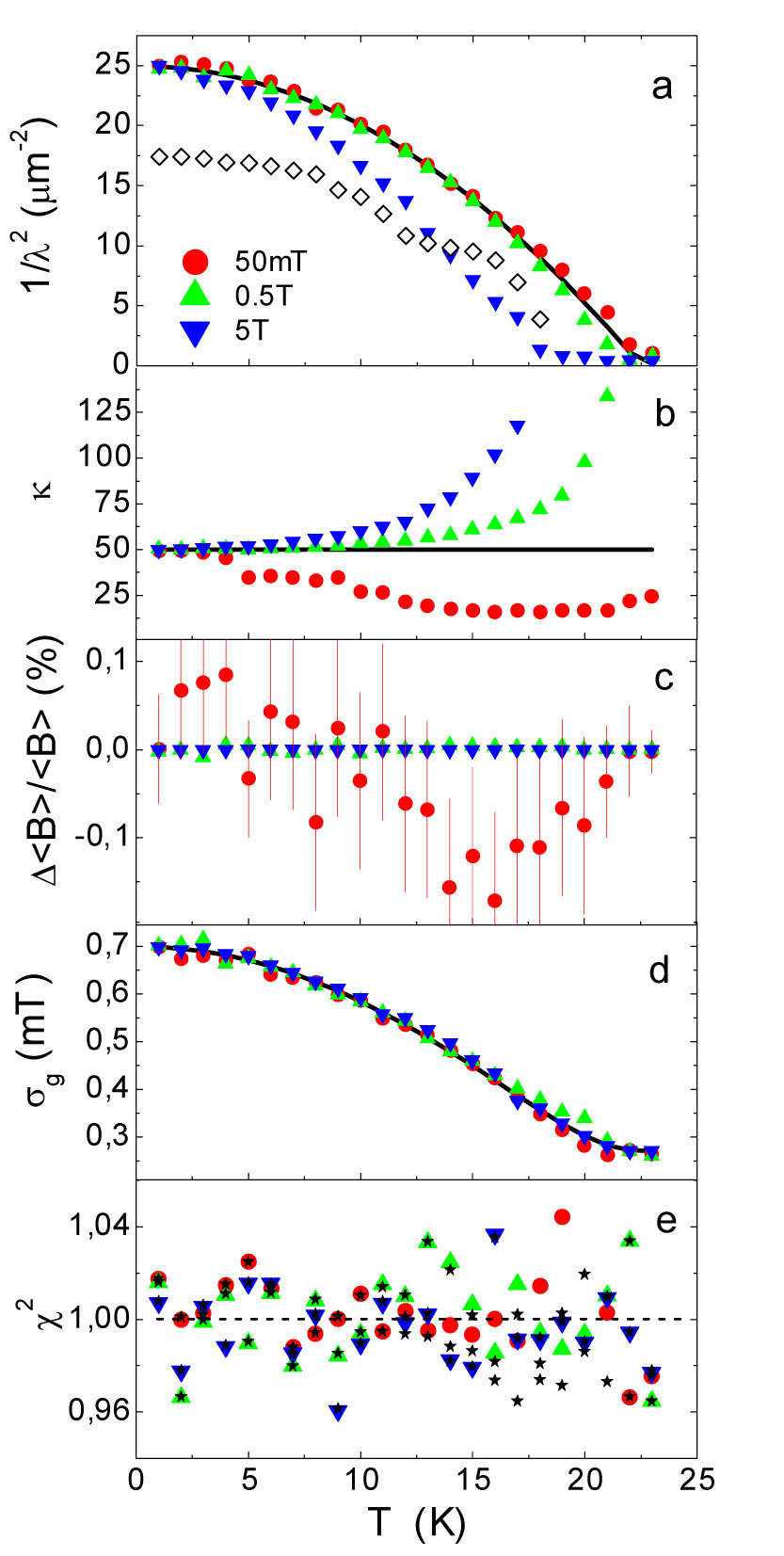}}
\caption{ Summary of the fit results for the $\mu$SR spectra simulated by the LG model. The fitted values $1/\lambda^2$, $\kappa$, $\Delta \langle B\rangle /\langle B\rangle$, and $\sigma_g$, as well as $\chi^2$ are plotted as a function of temperature $T$ for three different fields 0.05~T (circles), 0.5~T (up-triangles), and 5~T (down-triangles) for 20 million statistics. The parameters $1/\lambda^2$, $\kappa$, $\Delta \langle B\rangle /\langle B\rangle$, and $\sigma_g$ were free during the fitting procedure. The solid lines correspond to the true values of the parameters. $\Delta \langle B \rangle/\langle B \rangle = (\langle B \rangle_{fit} - \langle B \rangle_{real})/\langle B \rangle_{real}$ denotes the relative deviation of the fitted value $\langle B \rangle_{fit}$ from the real $\langle B \rangle_{real}$. For comparison $\chi^2$ was also determined for the real values of the parameters (black stars).  The black empty diamonds in (a) show a possible fit result for $1/\lambda^2$ at 5~T with extremely wrong initial parameters. Note that the error bars of  $1/\lambda^2$, as calculated by the fitting program (the function "fit" of the MATLAB program was used), are within the point size ($\simeq 2\%$). }\label{fig:FitFree}
\end{figure}

\begin{figure}[!htb]
\centering
\center{\includegraphics[width=0.45\textwidth]{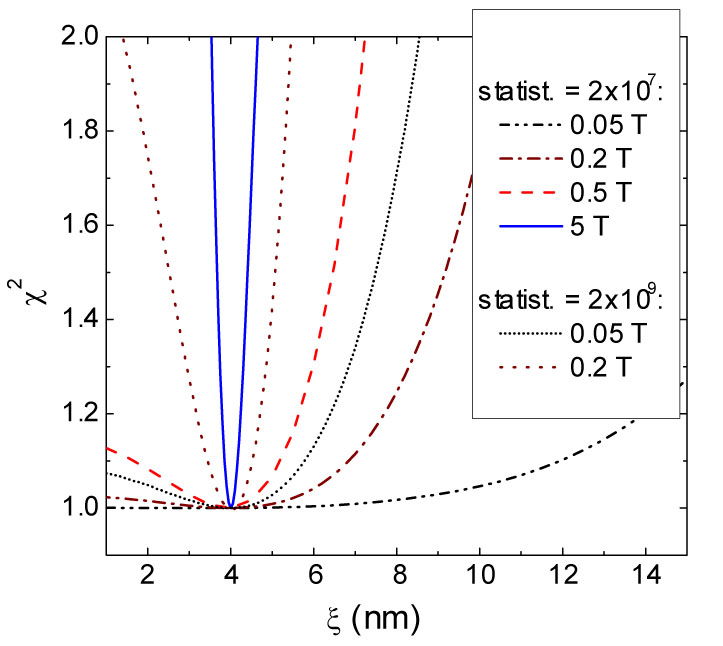}}
\caption{ $\chi^2$ as a function of $\xi$ for the $\mu$SR spectra simulated by the LG model with 20 million counts, $\langle B \rangle = 0.05$, 0.2,  0.5, and 5~T, $\lambda = 200$~nm, $\xi = 4$~nm, and $\sigma_g = 0.7$~mT. For comparison $\chi^2(\xi)$ is also shown for the higher statistics of  2000~millions at $\langle B \rangle = 0.05$ and 0.2~T. The statistically scattered minimal value of $\chi^2$ was normalized to 1. Note that the  dependence of $\chi^2$ on $\xi$ at low fields is weak. }\label{fig:ChiSqVsXi}
\end{figure}

In order to test the reliability of the advanced methods described in Sec.~II we first simulated noisy $\mu$SR spectra and then fitted them in a similar way as for the second moment method. To avoid systematic errors in the fit results it is important to analyze the data with the same model as they were simulated. Here we present results to the LG model, since it can well approximate experimental data in the whole field range (see Fig.~\ref{fig:ModelComp}).  Similar results are also obtained with all other models.    The temperature dependence of the penetration depth $\lambda$ was assumed to follow the relation $\lambda^{-2}(T)/\lambda^{-2}(0) = [1 - (T/T_c)^2 ]$ with $\lambda(0) = 200$~nm and $T_c = 22.5$~K. The Ginzburg-Landau parameter $\kappa = 50$, and Gaussian smearing $\sigma_g(T) = (\sigma_{VD}^2(T) + \sigma_N^2)^{1/2}$ [$\sigma_{VD} \propto 1/\lambda^2(T)$] was chosen to be the same as in Sec.~\ref{sec:TestOfSMmethod}. Again the $\mu$SR spectra were simulated for three different mean fields $\langle B \rangle = 0.05$, 0.5, and 5~T ($b = 0.0024$, 0.024, 0.24). As before the mean field was assumed to be  temperature independent. Statistics, asymmetry, and phase of the $\mu$SR time spectra were chosen to be $20\cdot10^6$, 0.2, and 0, respectively.

\begin{figure}[!htb]
\centering
\center{\includegraphics[width=0.45\textwidth]{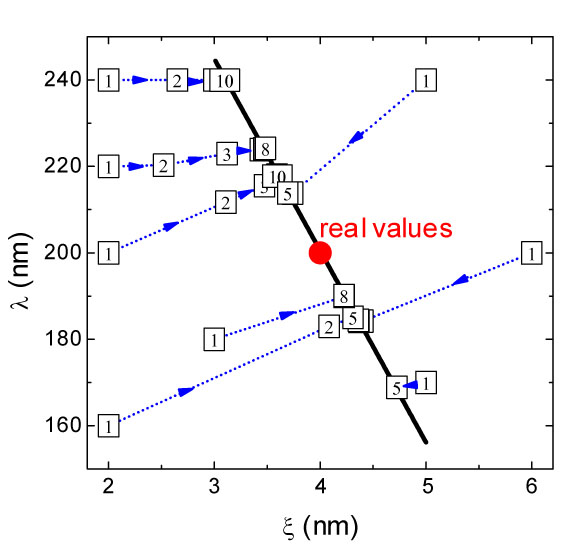}}
\caption{ Visualization of the fitting process of the simulated $\mu$SR spectrum for the parameters $\lambda = 200$~nm, $\xi = 4$~nm, $\langle B\rangle = 5$~T, $\sigma_g = 0.7$~mT using the LG model. Number \fbox{1} indicates the starting values for $\lambda$ and $\xi$. The numbers \fbox{2}, \fbox{3}, ...\fbox{10} denote the values of $\lambda$ and $\xi$ after each 5 iterations of the fitting process. The fit was terminated when relative changes of all the parameters were less than $10^{-6}$. The solid line shows the points in the $\lambda$ vs. $\xi$-plane where the fit is finally converging. Note that the fit results for the other parameters $\langle B\rangle$ and $\sigma_g$ are close to the real ones, independent of the starting values.   }\label{fig:LGM_FitConvergence}
\end{figure}

The results of the fits of the simulated $\mu$SR spectra with all the parameters free (except $A$ and $\phi$) are shown in Fig.~\ref{fig:FitFree}. Phase $\phi$ and asymmetry  $A$ were assumed to be known and were fixed to their real values. In order to exclude any artificial influence on the fitting procedure we performed the fits in automatic mode. This means that with increasing temperature the fit results for temperature $T_i$ were used as initial parameters for the next temperature $T_{i+1}$. For $T_1$ (lowest temperature) the correct initial parameters were used.  As shown in Fig.~\ref{fig:FitFree} for $\lambda^{-2}$ we got good results at low fields, however there are substantial systematic deviations for $\kappa$. At the highest field there are substantial systematic deviations of the fitted values for both $\lambda$ and $\kappa$, although the goodness of fit $\chi^2$ for these fits is comparable to those for the correct parameters (Fig.~\ref{fig:FitFree}). The values of $\chi^2$ weighted and normalized to the degrees of freedom ($\approx$~8000) scatter around 1 within 0.04, as expected for the present degrees of freedom. The black stars in Fig.~\ref{fig:FitFree} denote $\chi^2$ for the true values of the parameters for comparison. From all the fitted parameters, the values obtained for $\kappa$ deviate mostly from the real ones for all the fields. In order to check the reason for this strong deviation  for $\kappa$ at 0.05~T, the fit was performed using different initial parameters. For $\lambda$, $\langle B \rangle$, and $\sigma_g$ correct values within a few percent were obtained, whereas  $\xi$ was found to be in the range 2-13~nm with a very good value of $\chi^2$. The dependence of $\chi^2$ on $\xi$ for different fields and statistics is demonstrated in Fig.~\ref{fig:ChiSqVsXi}. It is evident that at low fields and 20 million statistics the quality of the fit is practically independent of $\xi$ in a very broad range.   
For the 5~T data not only $\kappa$ but also $\lambda$ substantially deviates from the real value (see Fig.~\ref{fig:FitFree}). The good agreement at low temperatures is misleading, since it is only due to the correct initial parameters we set for the lowest temperature. The empty black diamonds in Fig.~\ref{fig:FitFree}(a) shows the fit result for intentionally extremely wrong starting parameters. In real measurements one never knows the optimal starting parameters. We performed fits of the 5~T data at $T=1$~K with different starting values of $\lambda$, $\xi$, $\langle B\rangle$, and $\sigma_g$. It was found that for $\langle B\rangle$ and $\sigma_g$ one obtains values close to the real ones, however, for $\lambda$ and $\xi$ this is not the case. Figure~\ref{fig:LGM_FitConvergence} shows the variation of the values of $\lambda$ and $\xi$ during the fitting process. The starting values of $\lambda$ and $\xi$ are indicated by number \fbox{1}. The numbers \fbox{2}, \fbox{3}, ... \fbox{10} indicate the values of  $\lambda$ and $\xi$ after each 5 fitting iterations. The maximal number of fitting iterations was not restricted. However, 50 iterations were usually sufficient, and the fit was terminated when the relative changes of all the parameters during the iteration were less than $10^{-6}$. The final results of the fit eventually correspond to local minima of $\chi^2$.  The fit converges on a certain $\lambda = \lambda(\xi)$ curve denoted by the black line in Fig.~\ref{fig:LGM_FitConvergence}, indicating a possible correlation between $\lambda$ and $\xi$ at high fields.
Therefore, we can conclude, that the determination of a reliable value of $\xi$ is problematic at low fields as expected (see Fig.~\ref{fig:XsiDepChFlds}). At higher fields not only the value of $\xi$, but also the value of $\lambda$ may systematically deviate from the real value. However, at low fields reliable values of $\langle B \rangle$, $\sigma_g$, and $\lambda$ may presumably be well determined from fits with the advanced models.

\begin{figure}[!htb]
\centering
\center{\includegraphics[width=0.45\textwidth]{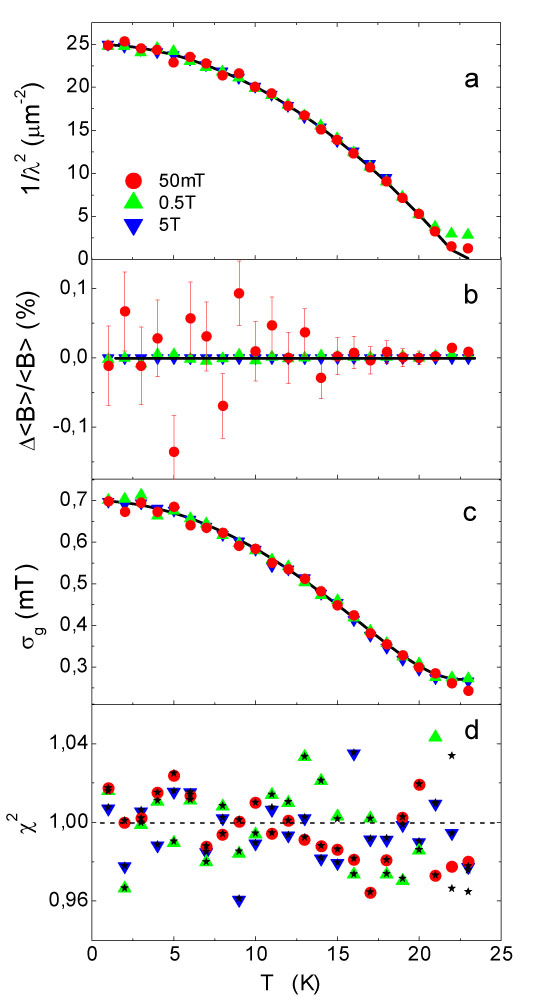}}
\caption{ Summary of the fit results for the $\mu$SR spectra simulated by the LG model. The fitted values $1/\lambda^2$, $\Delta \langle B\rangle /\langle B\rangle$, and $\sigma_g$, as well as $\chi^2$ are plotted as a function of temperature $T$ for three different fields 0.05~T (circles), 0.5~T (up-triangles), and 5~T (down-triangles) for 20 million statistics. The parameters $1/\lambda^2$, $\Delta \langle B\rangle /\langle B\rangle$, and $\sigma_g$ were free during the fitting procedure, whereas $\kappa$, was fixed at the real value. The solid lines correspond to the true values of the parameters. $\Delta \langle B \rangle / \langle B\rangle = (\langle B \rangle_{fit} - \langle B \rangle_{real})/ \langle B \rangle_{real}$ describes  the relative deviation of the fitted value $\langle B \rangle_{fit}$ from the real one $\langle B \rangle_{real}$. For comparison $\chi^2$ was also determined for the real values of the parameters (black stars).}\label{fig:FitKappa}
\end{figure}

The next step for improving the fitting procedure is to restrict some parameters. Based on the results obtained from the free parameters fits we conclude that a good candidate for restriction is $\xi$, especially at low fields. One can fix $\kappa$ in order to relate $\xi$ to $\lambda$ via $\xi = \lambda/\kappa$ and let the other parameters $\lambda$, $\langle B \rangle$, and $\sigma_g$ free. This is also reasonable from a theoretical point of view. In the BSC approximation  $\kappa$ changes not substantially with temperature  in the weak coupling limit \cite{Rainer74}.  The results of the fits with the only restricted parameter $\kappa$ (i.e., $\xi$ was calculated with $\xi = \lambda/\kappa$) are shown in Fig.~\ref{fig:FitKappa}. It is obvious, that the fits are excellent and all the parameters are very close to the real ones with only small statistical scattering. 

\subsection{Correlation between $\sigma_g$ and $\lambda^{-2}$ for small values of $b$}

\begin{figure}[htb]
\center{\includegraphics[width=0.4\textwidth]{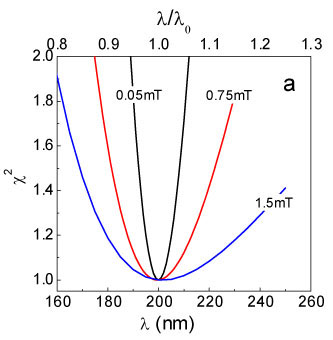}}
\center{\includegraphics[width=0.47\textwidth]{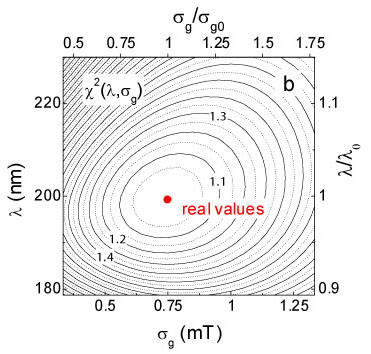}}
\caption{ (a) $\chi^2$ as a function of $\lambda$ with the other parameters set to the true values. The following parameters were used for the $\mu$SR data simulation: $\lambda_0 = 200$~nm, $\xi_0 = 1$~nm, $b_0 = 10^{-3}$, 20~millions statistics, and $\sigma_{g0}$ = 0.05, 0.75, and 1.5~mT. (b) Contour plot of $\chi^2$ as a function of $\lambda$ and $\sigma_g$ for the data simulated with $\lambda_0 = 200$~nm, $\xi_0 = 1$~nm, $b_0 = 10^{-3}$, $\sigma_{g0} = 0.75$, and 20 million counts (red point). In both figures the ML model was used for the calculations, and the statistically scattered minimal value of the $\chi^2$ was normalized to 1. The real values of $\lambda$ and $\xi$ are indicated by the red point. }\label{fig:CorrSigmaLambd}
\end{figure}

As shown in Fig.~\ref{fig:SigDepShape}, with increasing Gaussian smearing $\sigma_g$ the characteristic fields of the internal magnetic field distribution $P_{id}(B)$ of the ideal FLL are gradually washed out, and $P(B)$ tends to become an asymmetric Gaussian-like distribution. Therefore, one expects some correlation between $\sigma_g$ and the inverse square of the penetration depth $1/\lambda^2$, since both of them influence the second moment of the $\mu$SR spectrum. In order to show the possibility of extracting the real values of $\lambda$ and $\sigma_g$ from $\mu$SR spectra, we simulated $\mu$SR data by the ML model and then calculated the goodness of fit $\chi^2$ as a function of $\lambda$ and $\sigma_g$ with the other parameters fixed to their true values.   Figure \ref{fig:CorrSigmaLambd}(a) shows $\chi^2$ as a function of $\lambda$ with the other parameters fixed to their true values. For the simulated data the following parameters were used: $\lambda_0 = 200$~nm, $\xi_0 = 1$~nm, $b_0 = 10^{-3}$, $\sigma_{g0} = 0.05$, 0.75, 1.5~mT, and 20 million counts.              It is evident that with increasing $\sigma_{g0}$ the error of $\lambda$ extracted from the fit increases. Fig.~\ref{fig:CorrSigmaLambd}(b) shows a contour plot of $\chi^2$ as a function of $\lambda$ and $\sigma_g$ calculated for data with $\lambda_0 = 200$~nm, $\xi_0 = 1$~nm, $b_0 = 10^{-3}$, $\sigma_{g0} = 0.75$, and 20~millions statistics.  This approximately corresponds to the case we analyzed before. From the figure we conclude that $\lambda^{-2}$ and $\sigma_g$ are slightly correlated, but it is possible to extract both of them simultaneously if $\xi$ is fixed. This agrees well with the results of the analysis performed in Sec.~\ref{sec:TestOfLGm}.

\subsection{Correlation between $\xi$ and $\lambda$}\label{sec:corrLambdaXi}
\begin{figure}
\center{\includegraphics[width=0.45\textwidth]{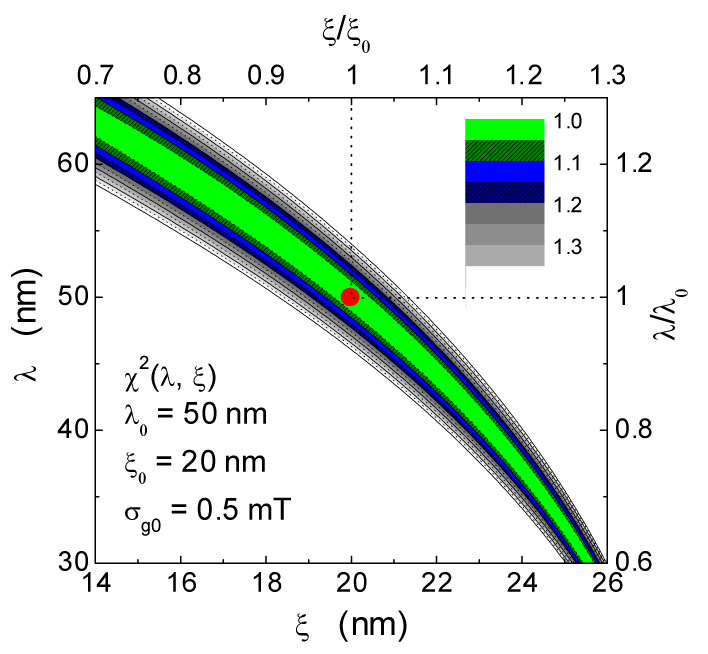}}
\caption{ Contour plot of $\chi^2$ as a function of $\lambda$ and $\xi$ for the data simulated with parameters $\lambda_0 = 50$~nm, $\xi_0 = 20$~nm, $b_0 = 0.5$, $\sigma_{g0} = 0.5$~mT, and 20 million counts,  calculated by the NGL model. The fitted parameters $\lambda$ and $\xi$ exhibit a strong correlation, which is the reason for the pronounced systematic deviations of the fitted values of $\lambda^{-2}$ and $\kappa$ at 5~T from the real values displayed in Fig.~\ref{fig:FitFree}. The statistically scattered minimal value of $\chi^2$ is normalized to 1. The real values of $\lambda$ and $\xi$ are indicated by the red point. }\label{fig:CorrLambdXi}
\end{figure}

For low magnetic fields the dependence of the $\mu$SR spectrum on the coherence length $\xi$ is very weak (see Fig.~\ref{fig:ChiSqVsXi}). But with increasing field towards $B_{c2}$ the  shape of the spectrum becomes dependent not only on the penetration depth $\lambda$, but also on $\xi$ as well (see Fig.~\ref{fig:ModelComp}). An increase of $\lambda$ and/or $\xi$  causes a decrease of the second moment and the characteristic fields. Therefore, it is expected that a decrease of $\lambda$  is correlated with an increase of $\xi$ in the fitting procedure and vice versa. So far, to our knowledge this problem was discussed previously only by Riseman \textit{et al.} \cite{Riesman95}.  Here we study this problem in more detail. For this procedure we determined $\chi^2$ for simulated $\mu$SR data as a function $\lambda$ and $\xi$ at fixed $\langle B \rangle_0 = 0.5B_{c2}$ and Gaussian smearing $\sigma_{g0} = 0.5$~mT.  We have chosen the case of a relatively small $\kappa_0 = 2.5$  ($\lambda_0 = 50$~nm and $\xi_0 = 20$~nm) for the data simulation, since the relative volume of the vortex cores is large, and therefore it is easier to extract $\xi$ from the fits. As before the statistics was $20\cdot 10^6$. The result of the analysis with the NGL model is shown in Fig.~\ref{fig:CorrLambdXi}, where a contour plot of $\chi^2$  as a function of $\lambda$ and $\xi$ with the other parameters fixed is displayed. There is indeed a strong correlation between $\lambda$ and $\xi$. For $\lambda(\xi) \approx 58.68 + 2.14\xi - 0.127\xi^2 $, where $\chi^2 \simeq 1$ is minimal, the fits converge after  a few hundreds of iterations. Tests showed that for different starting parameters the fits were converging in the correlated region of $\xi = 14-24$~nm and $\lambda \approx 36-63$~nm. This region lies within the region of $\chi^2 < 1.05$ (see Fig.~\ref{fig:CorrLambdXi}). It should be noted that for such a high reduced field $b = 0.5$ as used here for the analysis, the qualitative dependence of the characteristic fields $\delta B_{\alpha}$ on $\lambda$, $\xi$, and $\langle B\rangle$ are independent of $\kappa = \lambda/\xi$ (see Fig.~\ref{fig:ML_kappaDep}).  Therefore the qualitative behaviour of the contour plot of  $\chi^2(\lambda/\lambda_0, \xi/\xi_0)$ in Fig.~\ref{fig:CorrLambdXi} is independent of $\kappa_0 = \lambda_0/\xi_0$ for any $\kappa_0 > 5$.

\begin{figure}
\center{\includegraphics[width=0.45\textwidth]{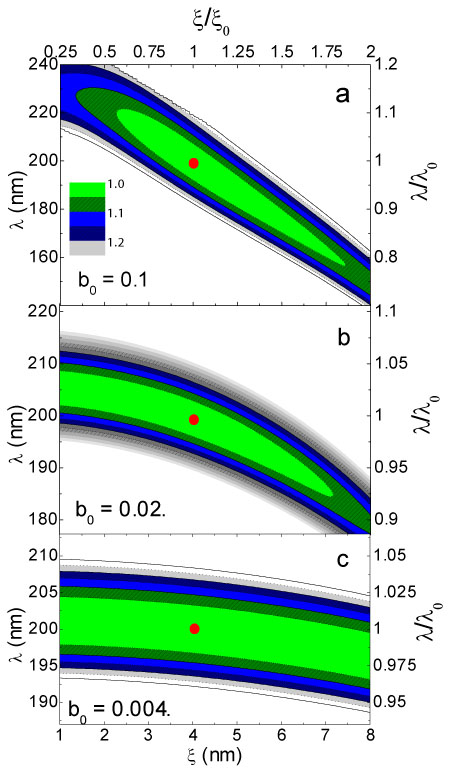}}
\caption{ Contour plots of $\chi^2$ as a function of  $\lambda$ and $\xi$ for $\mu$SR spectra simulated with the parameters $\lambda_0 = 200$~nm, $\xi_0 = 4$~nm, $\sigma_{g0} = 0.7$~mT, 20 million counts, and $b_0 = 0.004$, 0.02, and 0.1. The results in (a) were obtained by the NGL model, and those in (b) and (c) by the ML model. There is a strong correlation of the fitted values of $\lambda$ and $\xi$ for high ($b_0 = 0.1$) and intermediate ($b_0 = 0.02$) fields, but nearly no correlation at low fields ($b_0 = 0.004$). The statistically scattered minimal value of $\chi^2$ is normalized to 1. The real values of $\lambda$ and $\xi$ are indicated by the red point.}\label{fig:CorrLambdXi2}
\end{figure}

The situation could be improved for fields $b \simeq 10^{-2}$ where the minimal and saddle point fields slightly depend on $\xi$ (see Fig.~\ref{fig:ModelComp}), and on the other hand the maximal  field is not very large and still depends on $\xi$. However, with decreasing field, the vortex core volume substantially reduces which is disadvantageous for the data analysis as discussed above. Figure~\ref{fig:CorrLambdXi2} shows a similar contour plot of $\chi^2(\lambda, \xi)$ for noisy data simulated by the NGL and ML model with parameters $\lambda_0 = 200$~nm, $\xi_0 = 4$~nm ($\kappa_0 = 50$), $\sigma_{g0} = 0.7$~mT, 20 million statistics, and $b_0 = 0.004$, 0.02, and 0.1. This corresponds approximately to the analysis of the simulated data we discussed before. There is a substantial correlation between $\lambda$ and $\xi$ at all fields. For $b_0 = 0.1$ the analysis yields $\xi = 4(1)$~nm (correlated with $\lambda$), for $b_0 = 0.02$ only an upper limit of $\xi \simeq 5$~nm can be given, and for $b_0 = 0.004$ ($0.082$~T) the fit is practically independent of $\xi$ at 20 million statistics. However, at unrealistic high statistics the dependence of $\chi^2$ on $\lambda$ and $\xi$ becomes stronger (see Fig.~\ref{fig:ChiSqVsXi}), and the precision of the parameters extracted from the fit increases as the square root of the statistics. Another way of solving this problem was proposed by Riseman  \textit{et al.} \cite{Riesman95}, by simultaneously fitting several spectra measured at different fields with common values of the parameters $\lambda$ and $\xi$. This has two advantages: (1) It effectively increases statistics. (2) Since the correlation curve $\lambda(\xi)$ depends on field (see Fig.~\ref{fig:CorrLambdXi2}), the total contour graph of $\chi^2(\lambda, \xi)$ will shrink, allowing a determination of the correct parameters. For example in the case of a high value of $\kappa$ and extremely small field $b$,  one can determine the correct value of  $\lambda$ (independent of the value of $\xi$), and with the known value of $\lambda$ it is possible to evaluate a reliable value of $\xi$ at high field by means of the correlation curve $\xi(\lambda)$.
This procedure can be justified at least for conventional superconductors. Recently, Landau and Keller \cite{LandauKeller07} reanalyzed $\mu$SR data for various conventional superconductors and convincingly demonstrated, that in many cases type-II superconductors can be described by a field independent penetration depth.
The present results are also relevant for the interpretation of small-angle neutron scattering (SANS) experiments in the mixed phase of type-II superconductors, since the strong correlation between $\lambda$ and $\xi$ is also present in Fourier components of the FLL \cite{SANS06}.

We can conclude that in general a simultaneous determination of $\xi$ and $\lambda$ from $\mu$SR spectra without additional restrictions is not easy, independent of the model used to describe the vortex state. At high $\kappa$ and low fields there is practically no dependence of the spectra on $\xi$, and at high fields $\xi$ is strongly correlated with $\lambda$. As demonstrated in Figs.~\ref{fig:XsiDepChFlds},~\ref{fig:ML_kappaDep}, and~\ref{fig:ModelComp} this is independent of the value of $\kappa$ and the model used. It is important to add that in our analysis of the $\mu$SR data by the advanced models, we used the same model for the simulations and the analysis, what reduces systematic errors to a minimum.
In practice, there is often no adequate model for the description of the experimental $\mu$SR spectra as for instance for unconventional superconductors, such as the cuprate superconductors. In this case additional difficulties in the data analysis are expected.
In the analysis of the $\mu$SR spectra in section~5 we did not consider background signals
arising from impurity fractions/phases in the sample and/or from muons stopping in the sample holder or other parts of the spectrometer. These background signals may be a hidden source of uncertainties
in the determination of reliable parameters from $\mu$SR spectra.
The introduction of additional fit parameters in the advanced models should be done only with great care, since already the existing minimal set of parameters of the models are in general difficult to extract.

\section{Conclusions}

We performed an analysis of the line shape of $\mu$SR spectra of type-II superconductors in the mixed state simulated by four different models frequently used: (1) the modified London model (ML), (2) the London model with Gaussian cut-off (LG), (3) the analytical Ginzburg-Landau  (AGL), and (4) the numerical Ginzburg-Landau (NGL) model. The dependence of the line shape on the penetration depth $\lambda$, the coherence length $\xi$, the applied magnetic field $B$, and the Gaussian smearing parameter $\sigma_g$ is in agreement with previous studies \cite{BrandtJLTPhys88and77, Sidorenco90, Brandt03}. It is discussed under what conditions these models can be used to describe the vortex state in extreme type-II superconductors. As a result, the ML model can be applied for fields $b =  B/B_{c2} \leq 0.1$ ($B_{c2}$ is second critical field). On the other hand, the AGL and the LG model can be applied in the whole range of fields, but in the range $b \simeq 10^{-2} - 1$ they systematically deviate from the NGL model.  %
It was  shown that  at low fields $b \leq 10^{-3}$ there is practically no dependence of the line shape on $\xi$. However, with increasing field, there is a strong dependence of the line shape on both $\lambda$ and $\xi$, but the strong correlation between them makes it almost impossible to determine $\lambda$ and $\xi$ simultaneously. This is  independent of $\kappa = \lambda/\xi$ and the model used. Additional restrictions for $\xi$ (or $\lambda$) are needed to get rid of this correlation for reasonable statistics. Furthermore, it was shown  that it is possible to determine $\lambda$ and $\sigma_g$ simultaneously, provided that $\xi$ is fixed and the correlation between them is not too strong. In addition, it was demonstrated that the second moment method (SM), frequently used for $\mu$SR data analysis, may yield reliable values for $\lambda$ (within a few percent) in whole field range $0 < b \lesssim 1$, provided that vortex lattice disorder is not substantial. A multiple Gaussian fit may give reliable values for the second moment and may well approximate the local magnetic field distribution in a type-II superconductor. In order to substantiate these conclusions made above, we performed virtual experiments by generating noisy $\mu$SR spectra with known parameters. The results of a comprehensive analysis of these $\mu$SR spectra are in full agreement with the conclusions made above.

\begin{acknowledgments}

We would like to acknowledge E.H. Brandt and I.L. Landau for valuable discussions. This work was partly supported by the Swiss National Science Foundation, the K. Alex M\"uller Foundation, and the SCOPES GRANT No. IB7420-110784.
\end{acknowledgments}


%

\end{document}